%% file: main.tex
\documentclass{article}

\usepackage{amsfonts}
\usepackage{amssymb}
\usepackage{amsthm}
\usepackage{amsmath}
\usepackage{graphicx}
\usepackage{subcaption}
\usepackage{wrapfig}
\usepackage[final]{arxiv}

\usepackage{times}

\usepackage{tikz}
\usepackage{dsfont}
\usepackage{amsfonts}
\usepackage{multicol}
\usepackage{multirow}
\usepackage[shortlabels]{enumitem}
\usepackage{bm}
\usepackage{xr}

\setlist[enumerate,1]{leftmargin=0.5cm}
\setlist[itemize,1]{leftmargin=0.5cm}

\author{
    \textbf{Yuchen Cui}\textsuperscript{\rm 1}\thanks{Equally contributing authors}\space, 
    \textbf{\space Qiping Zhang}\textsuperscript{\rm 1}\footnotemark[1]\space,
    \textbf{\space Alessandro Allievi}\textsuperscript{\rm 1,2},\\
    \textbf{Peter Stone}\textsuperscript{\rm 1}\textbf{,}
    \textbf{\space Scott Niekum}\textsuperscript{\rm 1}\textbf{,}
    \textbf{\space W. Bradley Knox}\textsuperscript{\rm 1,2}  \\
    \textsuperscript{\rm 1}The University of Texas at Austin, Austin, TX\\
    \textsuperscript{\rm 2}Robert Bosch LLC, Austin, TX\\
    \{yuchencui, qpzhang\}@utexas.edu,
    \{pstone, sniekum\}@cs.utexas.edu \\
    \{brad.knox, alessandro.allievi\}@us.bosch.com
}

\title{The EMPATHIC Framework for Task Learning from Implicit Human Feedback
}

\begin{document}
\maketitle
\makeatletter{\renewcommand*{\@makefnmark}{}
	\footnotetext{Published as a conference paper at CoRL 2020. }
	\footnotetext{Project website: \url{sites.google.com/utexas.edu/empathic}}\makeatother}



\begin{abstract}
Reactions such as gestures, facial expressions, and vocalizations are an abundant, naturally occurring channel of information that humans provide during interactions.
A robot or other agent could leverage an understanding of such \emph{implicit} human feedback to improve its task performance at no cost to the human. This approach contrasts with common agent teaching methods based on demonstrations, critiques, or other guidance that need to be attentively and intentionally provided. In this paper, we first define the general problem of learning from implicit human feedback and then propose to address this problem through a novel data-driven framework, \textsc{empathic}. This two-stage method consists of (1) mapping implicit human feedback to relevant task statistics such as reward, optimality, and advantage; 
and (2) using such a mapping to learn a task. 
We instantiate the first stage and three second-stage evaluations of the learned mapping. To do so, we collect a dataset of human facial reactions while participants observe an agent execute a sub-optimal policy for a prescribed training task. 
We train a deep neural network on this data and demonstrate its ability to (1) infer relative reward ranking of events in the training task from prerecorded human facial reactions; (2) improve the policy of an agent in the training task using live human facial reactions; and (3) transfer to a novel domain in which it evaluates robot manipulation trajectories.

\end{abstract}


\input{sections/1.introduction.tex}

\input{sections/2.related_work.tex}
\input{sections/3.methodology.tex}
\input{sections/4.model_design.tex}
\input{sections/5.results.tex}

\input{sections/6.conclusion.tex}

\clearpage


\section{Broader Impacts}

In this work, we propose a data-driven framework for learning from implicit human feedback, enabling autonomous agents to leverage information that already exists during their interaction with end-users. Our proposed method and the collected dataset are part of an initial investigation of how to learn from implicit human feedback and therefore is not intended to be production-ready. Nevertheless, we identify the potential benefits and risks of our proposed method.

\textbf{Benefits} ~~ Our proposed method will benefit applications of autonomous learning agents that operate in human-centered environments. We envision \textsc{empathic} to be a complementary method for existing approaches that learn from explicit human feedback, when providing such explicit feedback is desirable. In particular, autonomous systems equipped with \textsc{empathic} will be able to interpret implicit human feedback as signals for learning tasks in which explicit human feedback signal is sparse or unavailable. Deployed systems will be able to adapt to their end-users' desires and preferences through incorporating implicit feedback using our proposed method, inducing little to no additional teaching cost (time and effort) for the end-user.

\textbf{Risks} ~~ The \textsc{empathic} framework makes use of implicit human feedback data and therefore shares many of the identified risks with machine learning applications that use personal data \cite{kamarinou2016machine,crawford2019ai,cave2019bridging}, including potential discrimination and breach of privacy. Here we focus on discussing potential risks that are introduced by misuse of \textsc{empathic}:
\begin{itemize}
	
	\item \textbf{Data Bias} ~~ \textsc{empathic} may lead to unexpected learning behavior when used to transfer learned reaction mappings to observers who are not sufficiently represented in the training data. \textsc{empathic} does not make the assumption that implicit reaction mappings fully generalize across individuals, and any particular learned model may work better for some populations over others, based on those populations' representations in the training set. Furthermore, the mappings should not be interpreted as describing any intrinsic characteristics of a person, but only as an interpretation of contextual implicit feedback that they are providing.
	
	\item \textbf{Non-consensual Use of Data} ~~ \textsc{empathic} could be misused in applications that capture a person's reactions to stimuli without their consent---for example, observing their reactions to advertisements, political messaging, or online content placement---to infer their beliefs about sensitive topics, to improve the persuasiveness of messaging, or to influence the behavior of users.
	
	\item \textbf{Adversarial Reactions} ~~ \textsc{empathic} leverages implicit human feedback. However, if someone is aware that the system adapts to their reactions, they may intentionally change their behavior to manipulate the agent. Intentional manipulation of \textsc{empathic} could be harmful if the agent is deployed amongst other people, such as a robot in a hospital that attempts to navigate busy walkways without causing disruption.

\end{itemize}

\acknowledgments{
	
	Part of this work has taken place in the Personal Autonomous Robotics Lab (PeARL) at The University of Texas at Austin. PeARL research is supported in part by the NSF (IIS-1724157, IIS-1638107, IIS-1749204, IIS-1925082) and ONR (N00014-18-2243).  This research was also sponsored by the Army Research Office and was accomplished under Cooperative Agreement Number W911NF-19-2-0333. The views and conclusions contained in this document are those of the authors and should not be interpreted as representing the official policies, either expressed or implied, of the Army Research Office or the U.S. Government. The U.S. Government is authorized to reproduce and distribute reprints for Government purposes notwithstanding any copyright notation herein.
	
	A portion of this work has taken place in the Learning Agents Research
	Group (LARG) at UT Austin.  LARG research is supported in part by NSF
	(CPS-1739964, IIS-1724157, NRI-1925082), ONR (N00014-18-2243), FLI
	(RFP2-000), ARO (W911NF-19-2-0333), DARPA, Lockheed Martin, GM, and
	Bosch.  Peter Stone serves as the Executive Director of Sony AI America
	and receives financial compensation for this work.  The terms of this
	arrangement have been reviewed and approved by the University of Texas
	at Austin in accordance with its policy on objectivity in research.
	
}

\clearpage
\bibliography{references}
\clearpage

\appendix
\normalsize \input{appendix.tex}

\end{document}

%% file: sections/1.introduction.tex
\section{Introduction}
\label{sec:introduction}


People often react when observing an agent---whether human or artificial---if they are interested in the outcome of the agent's behavior. We have scowled at robot vacuums, raised eyebrows at cruise control, and rebuked automatic doors. 
Such reactions are often not intended to communicate to the agent and yet nonetheless contain information about the perceived quality of the agent's performance. A robot or other software agent that can sense and correctly interpret these reactions could use the information they contain to improve its learning of the task. 
Importantly, learning from such \emph{implicit} human feedback does not burden the human, who naturally provides such reactions even when learning does not occur. We view learning from implicit human feedback (\textsc{lihf}) as complementary to learning from explicit human teaching, which might take the form of demonstrations \cite{argall2009survey}, evaluative feedback \cite{knox2009interactively,knox2013training}, or other communicative modalities \cite{chernova2014robot,sadigh2017active,cui2018active,kroemer2019review,admoni2017social}. 
Though we expect implicit feedback to typically be less informative in a fixed amount of time than explicit alternatives and perhaps more difficult to interpret correctly, \textsc{lihf} has the advantage of using already available reactions and therefore induces no additional cost to the user.
The goal of this work is to frame the \textsc{lihf} problem, propose a broad data-driven framework to solve it, and implement and validate an instantiation of this framework using specific modalities of human reactions: facial expressions and head poses (henceforth referred to jointly as \textbf{facial reactions}).


Existing computer vision research has shown success in recognizing basic human facial expressions \cite{ekman1999facial, fasel2003automatic, li2018deep}. However, it is not trivial for a learning agent to interpret human expressions. For example, a smile could mean satisfaction, encouragement, amusement, or frustration \cite{hoque2012exploring}.
Different interpretation of the same facial expression could result in very different learning behaviors. Recent progress in cognitive science also provides a utilitarian view of facial expressions and suggests that they are also used as tools for regulating social interactions and signaling contingent social action; therefore the interpretation of facial expressions may vary from context to context and from person to person \cite{panksepp2011basic, crivelli2018facial,jack2015human,dailey2010evidence}.
Further, human reactions often have a variable delay after an event or occur in anticipation of an event, posing an additional challenge of interpreting which (series of) action(s) or event(s) the person is reacting to. 
Additionally, many natural human reactions involve spontaneous micro-expressions consisting of minor facial muscle movements that last for less than 500 milliseconds \cite{pfister2011recognising,yan2013fast}, which can be hard to detect by computer vision systems trained with common datasets with only exaggerated or acted facial expressions~\cite{li2013spontaneous,davison2018objective}. 
Lastly, human environments often contain more than the agent and its task environment, and therefore inferring what a person is reacting to at any moment adds further difficulty. 

We approach \textsc{lihf} with data-driven modeling that creates a general \textbf{reaction mapping} from implicit human feedback to task statistics. 
The major contributions of this paper are:

\begin{enumerate}
    \item We motivate and frame the general problem of Learning from Implicit Human Feedback (\textsc{lihf}), which aims at leveraging under-utilized data modality that already exists in natural human-robot interactions. This problem is different from traditional interactive robot learning settings that put human and robot in explicit pedagogical settings.
    
    \item We propose a general framework to solve this problem, called Evaluative MaPping for Affective Task-learning via Human Implicit Cues (\textsc{empathic}), which consists of two stages: (1) learning a mapping from implicit human feedback to known task statistics and (2) using such a mapping to learn a task from implicit human feedback. 
   
    \item We experimentally validate an instantiation of the \textsc{empathic} framework, using human facial reactions as the implicit feedback modality, and rewards as target task statistic:
    
    \begin{itemize}
    
    \item We develop an experimental procedure for collecting data of human reactions to an autonomous agent's behavior. The dataset is recorded while \textit{human observers} watch an autonomous agent performing a task that determines their financial payout. We refer to such tasks as the \textbf{training tasks}.
    
    \item We analyze the modeling problem through a \textit{human proxy test}: the authors act as proxies for a reaction mapping by watching the reactions of the human observers and then ranking semantically anonymized events by their inferred reward, which we refer to as the \textit{reward-ranking task}. Moderate success at this human proxy test provides confidence that human reactions could inform an understanding of reward values. This activity also provides critical insight regarding which reaction features are helpful for modeling.
    
    \item Our instantiation of \textsc{empathic} learns a \textbf{reaction mapping} from a proximate time window of human reactions to a probability distribution over reward values. The mapping is learned by using a
    pre-trained model to extract facial reaction features from video data and training a deep neural network via supervision to predict rewards with the extracted features. 
    
    \item We compare the performance of the learned reaction mapping and a random baseline on the reward-ranking task. We also show an initial evaluation of learning the training task \textit{online}, in which an agent updates its belief over possible reward functions from live human reactions and improves its policy in real time. 

    \item We transfer the learned reaction mapping to a \textbf{deployment task}, providing a proof-of-concept of the potential for reaction mappings to generalize across tasks. 
    Specifically, the reaction mapping trained with data from the training task is used to evaluate and rank trajectories from a robotic sorting task. 
    \end{itemize}

\end{enumerate}

%% file: sections/2.related_work.tex
\section{Related Work}
\label{sec:related_work}

Our work relates closely to the growing literature of \textit{interactive reinforcement learning (RL)}, or human-centered RL \cite{knox2009interactively, isbell2001social, pilarski2011online, suay2011effect, warnell2018deep, li2019human, zhang2019leveraging, lin2020review, li2020facial, macglashan2017interactive}
, in which agents learn from interactions with humans in addition to, or instead of, predefined \textit{environmental} rewards. 
In the \textsc{empathic} framework, we use the term \textit{implicit} human feedback to refer to any multi-modal evaluative signals humans naturally emit during social interactions, including facial expressions, tone of voice, head gestures, hand gestures, and other body-language and vocalization modalities not aimed at explicit communication. 
Others' usage of ``implicit feedback'' has referred to the \textit{implied} feedback when a human refrains from giving explicit feedback \cite{loftin2014learning, joachims2017accurately}, to human biomagnetic signals \cite{xuplaying}, or to facial expressions \cite{jaques2018learning,arakawa2018dqn,veeriah2018beyond}.
This work focuses on predicting task statistics from human facial features and therefore is also related to the broad area of research on \textit{facial expression recognition}.

\textbf{Interactive Reinforcement Learning}~
Inspired by clicker-training for animals, the \textsc{tamer} framework proposed by Knox et al.~\cite{knox2009interactively, knox2013training} is the first to explicitly model human feedback in the form of button clicks, thus allowing RL agents to learn from human feedback signals without any access to environmental rewards. 
\citet{veeriah2016face} propose learning a value function grounded only in the user's facial expressions and agent actions, using manual negative feedback as supervision. 
The corresponding RL agent's policy is only a function of the trainer's facial expression and does not reason about the task state.
In the preliminary work of \citet{arakawa2018dqn}, the authors adopt an existing facial expression classification system to detect human emotions and use a predefined mapping from emotions to \textsc{tamer} feedback but do not optimize the mapping to be effective for the downstream task. Similarly, recent work of \citet{zadok2019affect} models the probability of human demonstrators smiling within a task and then biases an RL agent's behavior to increase the predicted probability of human smiling, improving exploration. 
\citet{li2020facial} extend \textsc{tamer} by interpreting the trainer's facial expressions as positive or negative feedback with a deep neural network. Their results suggest it is possible to learn solely from facial expressions of the trainer.
Our proposed method differs from prior work through learning a direct mapping from facial reactions to task statistics independent of states or state-actions, which requires no explicit human feedback at either training or testing time. Our system is the first, to the best of our knowledge, attempting to learn from subjects that are not explicitly told to teach or react. 

\textbf{Facial Expression Recognition (\textsc{fer})}~
The field of facial expression recognition contains a rich body of research from
areas of psychology, neuroscience, cognitive science and computer vision. \citet{fasel2003automatic} provide an overview of traditional \textsc{fer} systems and \citet{li2018deep} detail recent \textsc{fer} systems based on deep neural networks.
Our proposed method does not perform \textsc{fer} explicitly but maps extracted facial features to reward values. 
Our work is closely related to the problem of dynamic \textsc{fer}, where time-series data are used as input for temporal predictions. Modern \textsc{fer} systems often consist of two stages: data pre-processing and predictive modeling with deep networks~\cite{li2018deep}. 
Inspired by techniques from the \textsc{fer} literature, our proposed system leverages an existing toolkit  \cite{baltrusaitis2018openface, baltruvsaitis2015cross,zadeh2017convolutional} to extract facial features that are sufficiently informative for modeling despite our small dataset, and we explicitly model the temporal aspect of the problem by further extracting features in the frequency domain. 

%% file: sections/3.methodology.tex
\vspace{-0.04cm}
\section{The LIHF Problem and The EMPATHIC Framework}
\label{sec:methodology}

\textbf{Markov Decision Processes} (MDPs) are often used to model sequential decision making of autonomous agents. An MDP is given by the tuple  $\langle S,A,T,R,d_0,\gamma \rangle$, where:
$S$ is a set of states; $A$ is a set of actions an agent can take; 
$T: S \times A \times S \to [0,1] $ is a probability function describing state transition based on actions;
$R: S \times A \times S \to \mathbb{R}$ is a real-valued reward function;
$d_0$ is a starting state distribution and $\gamma \in [0,1)$ is the discount factor.
A policy $\pi: S \times A \to [0,1]$ maps from any state and action to a probability of choosing that action. The goal of an agent is to find a policy that maximizes the expected return $E\left[\sum_{t=0}^{\infty} \gamma^{t} r_t\right]$ where $r_t$ is the reward at time step $t$.

The problem of \textbf{Learning from Implicit Human Feedback} (\textsc{lihf}) asks how an agent can learn a task with information derived from human reactions to its behavior.
\textsc{lihf} can be described by the tuple $\langle S,A,T,R^{\mathcal{H}},X^{\mathcal{H}},\Xi,d_0,\gamma \rangle$.  
$S,A,T,d_0,$ and $\gamma$ are defined identically as in MDPs. 
The agent receives observations from implicit feedback modalities asynchronously with respect to time steps, and each such observation $x\in X^{\mathcal{H}}$ contains implicit feedback from some human $\mathcal{H}$. 
An observation function $\Xi$ denotes the conditional probability over $X^{\mathcal{H}}$ of observing $x$, given a trajectory of states and actions and the human's hidden reward function $R^\mathcal{H}$. States in \textsc{lihf} are generally broader than task states, and include all environmental and human factors that influence the conditional probability of observing $x$.
The goal of an agent is to maximize the return under $R^\mathcal{H}$. How to ground observations $x \in X^{\mathcal{H}}$ containing implicit human feedback 
to evaluative task statistics is at the core of solving \textsc{lihf}.

The formulation of \textsc{lihf} resembles the definition of Partially Observable MDPs, but here the partially observable variable is the human's reward function rather than state. We include a graphical model in Appx.\ref{Formulation} that describes how \textsc{lihf} models the data generation process. 

\begin{figure*}[t]
    \centering
    \includegraphics[width=\textwidth]{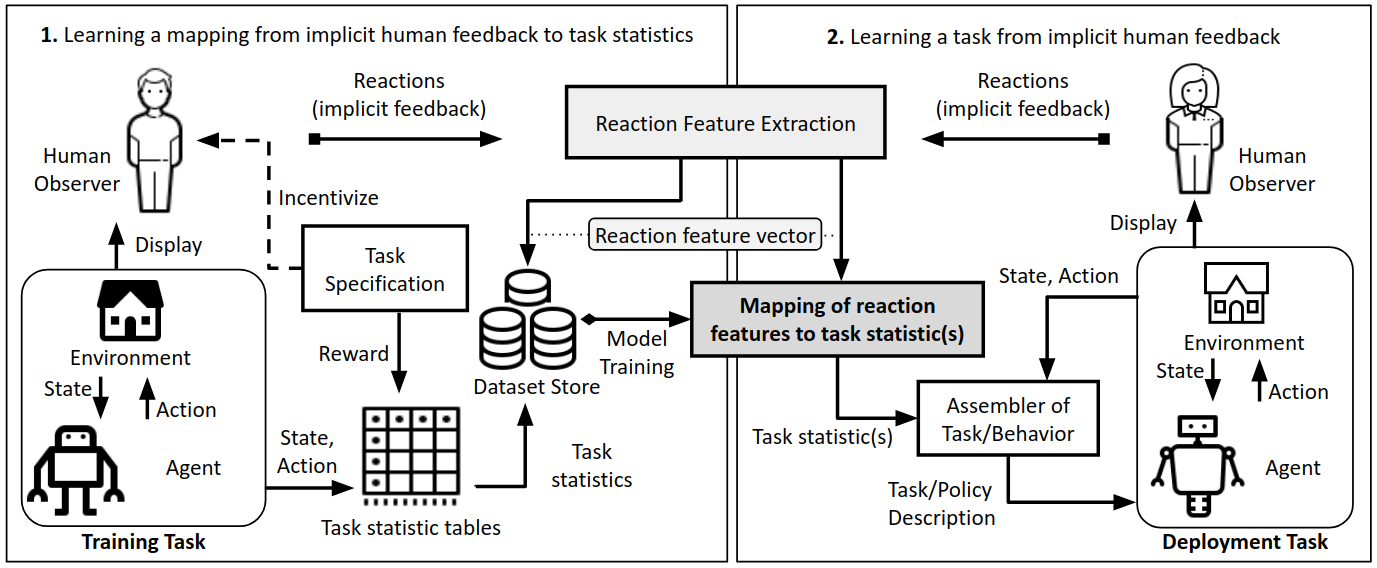}
    \vspace{-0.5cm}
    \caption{Overview of \textsc{empathic}}
    \label{fig:overview}
    \vspace{-0.6cm}
\end{figure*} 

We propose a data-driven solution to the \textsc{lihf} problem that infers relevant task statistics from human reactions.
As Fig.~\ref{fig:overview} shows, the \textsc{empathic} framework has two stages: (1) learning a mapping from implicit human feedback to relevant task statistics and (2) using such a mapping to learn a task. 
In the first stage, human observers are incentivized to want an agent to succeed---to align the person's $R^\mathcal{H}$ with a known task reward function $R$---and they are then recorded while observing the agent.
Task statistics are computed from $R$ for every timestep to serve as supervisory labels, which train a mapping from synchronized recordings of the human observers to these statistics. Task state and action are \textit{not} inputs to the reaction mapping, allowing it to be deployed to other tasks. In the second stage, a human observes an agent attempt a task with sparse or no environmental reward, and the human observer's reaction to its behavior is mapped to otherwise unknown task statistics to improve the agent's policy, either directly or through other usage of the task statistics, such as guiding exploration or inferring the reward function $R^\mathcal{H}$ that describes the human's utility.
In general, any instantiation of \textsc{empathic} can be achieved through specifying these elements: 
\begin{itemize}[leftmargin=.3in,noitemsep,nolistsep]
\vspace{-0.08cm}
\item
the reaction modality and the target task statistic(s);  ~~~~$\bullet$ the end-user population(s);
\item training task(s) for stage 1 and deployment task(s) for stage 2; 
\item an incentive structure for stage 1 to align human interests with task performance; and
\item policies or RL algorithms to control the observed agent in both stages. 
\end{itemize}
\vspace{-0.08cm}
Any specific task or person can optionally be part of both stages. 
Note that \textsc{empathic} is defined broadly enough to include instantiations with varying degrees of personalization---from learning a single reaction mapping applicable to all humans to training a person-specific model---and of across-task generalization. We hypothesize that a single reaction mapping will be generally useful but that training to specific users or tasks will yield even more effective mappings. Such personalized training may also guard against negative effects of potential dataset bias from the first stage of \textsc{empathic} if it is used amongst underserved populations. 



This paper presents one instantiation of \textsc{empathic}, using facial reactions as the modality for implicit human feedback. Sections~\ref{sec:data_collection} and \ref{sec:model_design} provide the instantiation details.


\begin{wrapfigure}{r}{.39\textwidth}
	\vspace{-1.6cm}
	\centering\captionsetup[subfigure]{justification=centering}
	\hspace{-0.15cm}\includegraphics[width=1.0\linewidth]{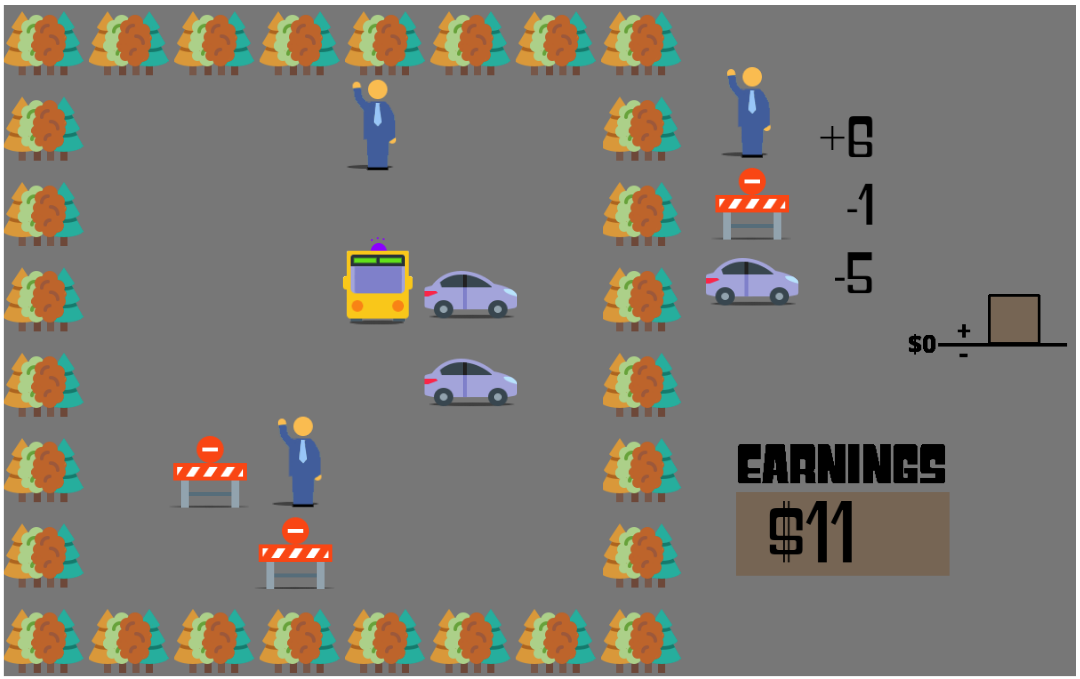}
	\vspace{-0.09cm}
	\caption{\small \textit{Robotaxi} environment}
	\label{fig:robotaxi_env}\par\vfill
	\vspace{-0.8cm}
\end{wrapfigure}

\section{Data Collection Domains and Protocol}
\label{sec:data_collection}

In this section we describe the experimental domains and data collection process of our instantiation of \textsc{empathic}.

\textbf{Robotaxi}~
We create \textit{Robotaxi} as a simulated domain to collect implicit human feedback data with known task statistics. 
Fig.~\ref{fig:robotaxi_env} shows the visualization viewed by the human observer.
An agent (depicted as a yellow bus) acts in a grid-based map. 
Rewards are connected to objects: $+6$ for picking up a passenger; 
$-1$ for crashing into a roadblock; and $-5$ for crashing into a parked car. Reward is $0$ otherwise. An object disappears after the agent moves onto it, 
and another 
object of 
the same type 
is spawned with a short delay at a random unoccupied location. An episode starts with two objects of each type.

\begin{wrapfigure}{r}{.28\textwidth}
	\vspace{-0.1cm}
	\centering\captionsetup[subfigure]{justification=centering}
	\includegraphics[width=0.99\linewidth,height=0.95\linewidth]{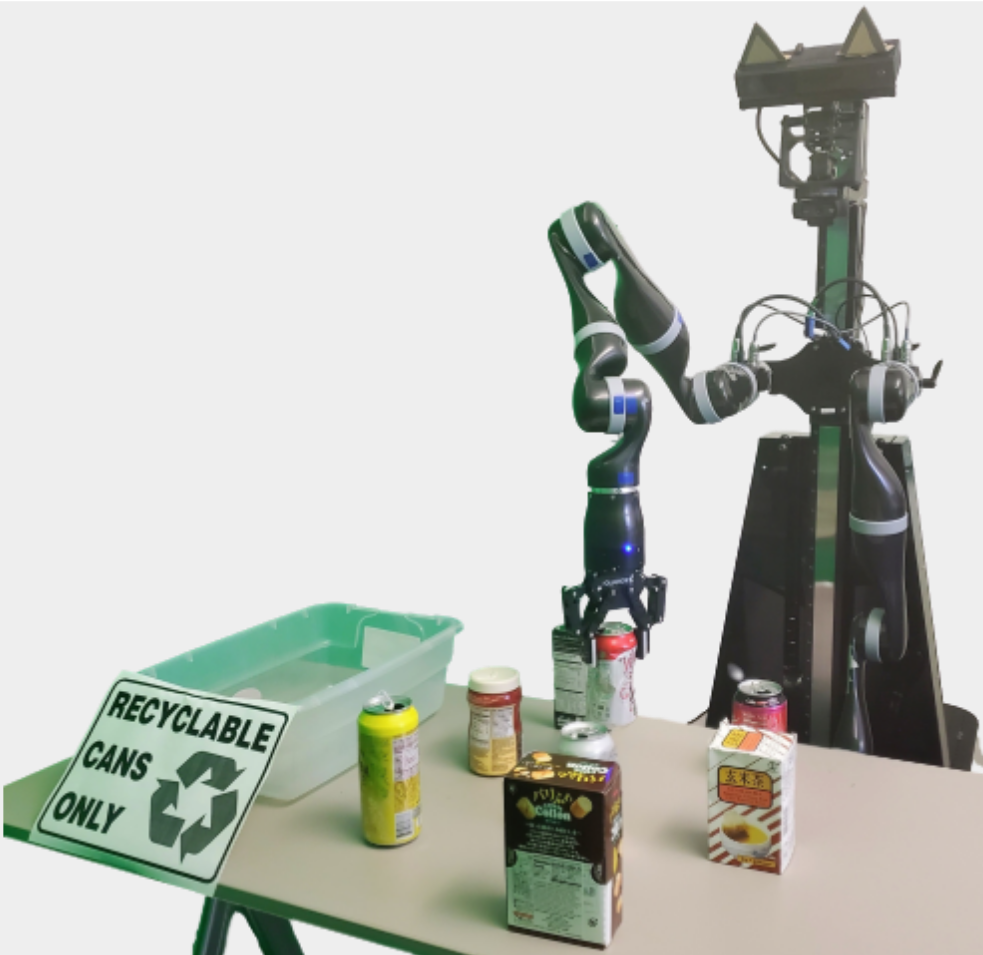}
	\vspace{-0.55cm}
	\caption{\small Robotic sorting task}
	\label{fig:robotic_task}
	\vspace{-0.5cm}
\end{wrapfigure}

\textbf{Robotic Sorting Task}~
A robotic manipulation task is a deployment domain for test transfer of the learned reaction mapping across task domains. The physical setup of the task is shown in Fig.~\ref{fig:robotic_task}. The robot's task is to sort the aluminum cans into the recycling bin. Reward is $+2$ upon recycling a can, $-1$ upon recycling any other object, and $0$ at all other times. The episodes are short ($<20$ seconds), containing predetermined trajectories with at most a single non-zero reward event. Further details are in Appx.\ref{Data Collection Details}.

\textbf{Data Collection}
We recruited participants to interact with autonomous agents in both tasks. Before human participants observed the agents executing a task, they were informed that their financial compensation for the study would be proportional to the agent's earnings. The payment structure creates a direct mapping between the ground-truth reward label and its financial value to the human subject, intending to align human interests with the task and therefore connecting their reactions to task statistics. 
To minimize explicit feedback (i.e., intended to influence the agent), participants were told that their ``reactions are being recorded for research purposes'', and nothing more was said regarding our intended usage of their reactions. 
This experimental setup contrasts with prior related work~\cite{li2020facial,arakawa2018dqn,veeriah2016face}, in which human participants were explicitly asked to teach with their facial expressions, 
and aligns with a key motivation for the \textsc{lihf} problem, which is to leverage data that is already being generated in existing human-agent interactions. 
17 human participants observed 3 episodes of \textit{Robotaxi}, and 14 of the participants observed 7 episodes of the robotic task. Experiments occurred in an isolated room and 
videos were recorded as the human participants watched the agents execute suboptimal behavior trajectories that were predefined. 
All data collection was conducted after obtaining the participant's consent and the participants were debriefed at the end of their sessions. 
See Appx.\ref{Data Collection Details} for further details.

%% file: sections/4.model_design.tex
\vspace{-0.1cm}
\section{Reaction Mapping Design}
\label{sec:model_design}
\vspace{-0.1cm}


\begin{table}[b]
\vspace{-0.5cm}
\begin{minipage}{0.635\textwidth}
        \vspace{-0.1cm}
       \centering
        \includegraphics[width=1.01\textwidth]{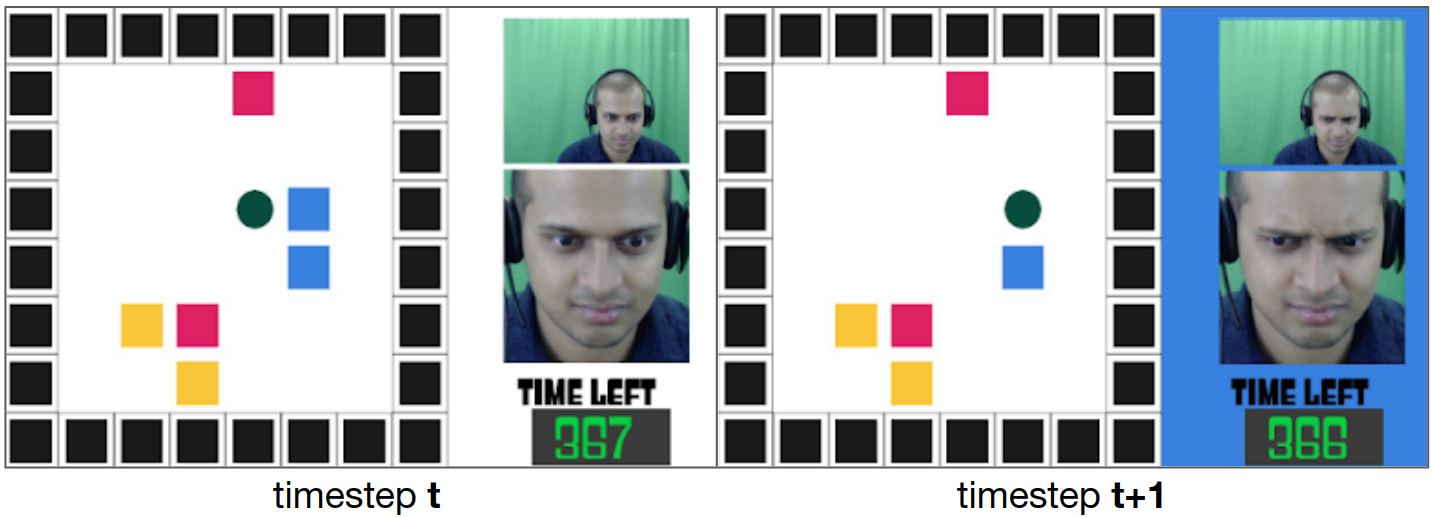}
        \vspace{-0.3cm}
        \captionof{figure}{\small Human proxy's view: semantics are hidden with color masks; the dark green circle is the agent; observer's reaction is displayed; detected face is enlarged; background is colored by last pickup. The left frame is the same game state shown in Fig~\ref{fig:robotaxi_env}.
        } 
        \label{fig:proxy_view}
        \vspace{-0.3cm}
\end{minipage}
~
\begin{minipage}{0.36\textwidth}
    \centering
    \small
    {\renewcommand{\arraystretch}{1.1}
    \begin{tabular}{|c|c|c|}
      \hline 
       	    & Avg. $\tau$ & p-value \\ \hline  
         & .569  & .004  \\ 
         & .216  & .185  \\ 
        Human & .098  & .319  \\ 
        Proxies & -.176 & .179 \\ 
         & .255  & .123  \\ 
         & .294  & .059  \\ \hline
        Avg.  & .209 & .078  \\   \hline 
    \end{tabular}}
     \vspace{0.15cm}
    \caption{\small Human proxy test result: average $\tau$ values across participants are displayed; a baseline that randomly picks rankings has an expected $\tau$ value of 0. } 
    \label{tab:proxy_test}
\end{minipage}

\end{table}

\textbf{Human Exploration of the Data} To better understand the task of training a reaction mapping, the authors serve as proxies for a mapping. Specifically, we view a semantically anonymized version of each agent trajectory alongside a synchronized recording of the human participant's reactions; after this viewing, we attempt to rank the reward values of the 3 object types. 
Fig.~\ref{fig:proxy_view} shows the interface.
Each human proxy watched one truncated episode from each of the 17 participants. To measure performance, Kendall's rank correlation coefficient $\tau \in [-1,1]$ \cite{abdi2007kendall} is used to compare a human proxy's inferred ranking with ground truth (a higher $\tau$ value indicates a higher correlation between two rankings). 
Table~\ref{tab:proxy_test} shows mean $\tau$ scores for the human proxy test across 17 participants, with a mean for each author. Wilcoxon signed-rank tests \cite{woolson2007wilcoxon} compare each human proxy's 17 $\tau$ scores with the expected value $\tau=0$ for uniformly random reward ranking, and corresponding p-values are also in Table~\ref{tab:proxy_test}.  
In this test, 5 out of 6 humans outperformed random ranking, and 1 human author did so significantly even after adjusting a $p<0.05$ threshold for multiple testing to $p<0.0083$ using a Bonferroni correction \cite{weisstein2004bonferroni}.
This person's success suggests that the reactions contain sufficient information to rank object rewards, though humans vary in their ability to harness the information. 
With our experience as proxies for the reaction mapping, we identify 7 common reaction gestures that helped us infer reward rankings: smile, pout, eyebrow-raise, eyebrow-frown, (vertical) head nod, head shake, and eye-roll. The collected video data was annotated with frame onsets and offsets of these 7 gestures as well as the general positive, negative, or neutral sentiment of the gesture. The corresponding trajectories were not viewed during annotation. Appx.\ref{Annotation Visualization} contains a detailed analysis of the annotations, which informed our model design.

\begin{figure*}[t]
    \centering
    \includegraphics[width=\textwidth]{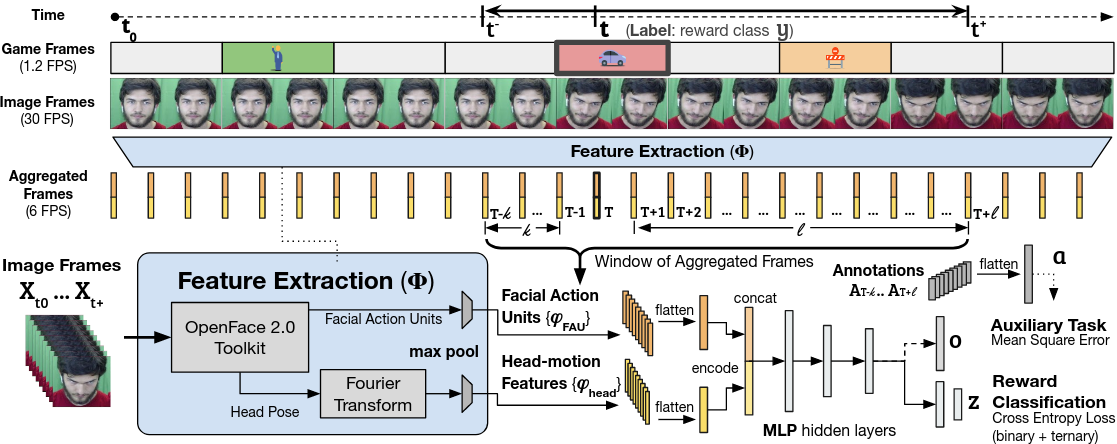}
    \vspace{-0.5cm}
    \caption{\small The feature extraction pipeline and architecture of the reaction mapping
    }
    \label{fig:model_architecture}
     \vspace{-0.57cm}
\end{figure*}

\setcounter{footnote}{0}
\textbf{Reaction Mapping Architecture} To demonstrate that the implicit feedback signal in human facial reactions can be computationally modeled, we construct a reaction mapping that takes a temporal series of extracted features as input and outputs a probability distribution over reward classes. We extract facial features from video data with a pre-trained model and train a deep neural network on predicting rewards with the extracted features in a supervised way.\footnote{Our proposed approach could instead model other task statistics or be trained end-to-end with a convolutional neural network (removing the feature extraction module). \textit{For this dataset}, however, modeling non-zero reward classes with a pre-trained feature extractor is empirically more effective than either of these strategies. More details can be found in Appx.\ref{Ablation} and \ref{RLstats}.}
The feature extraction pipeline and architecture of the proposed deep network model is shown in Fig.~\ref{fig:model_architecture}. OpenFace 2.0~\cite{baltrusaitis2018openface, baltruvsaitis2015cross,zadeh2017convolutional} is used to extract features from raw videos of human reactions. Raw videos consist of 30 image frames per second. 
For each image frame, OpenFace extracts head pose and activation of facial action units (FAUs).
For detecting head nods and shakes, we explicitly model the head-pose changes by keeping a running average of extracted head-pose features and subtract it from each incoming feature vector. Frequencies of changes in head-pose are then computed through fast Fourier transform, and the coefficients of frequencies are used as head-motion features.
To allow the series of input features to cover a large enough temporal window of reactions, 
feature vectors of consecutive image frames are combined through max pooling of each dimension, resulting in temporally aggregated feature vectors of the same size. 
Refer to Appx.\ref{Feature Extraction} for full details about feature extraction.

Let $\{X_{t_0},...,X_{t}\}$ denote the sequence of raw input image frames from time $t_0$ to $t$. Time $t_0$ is the start of an episode, and $t$ is the time of the last image frame for the $T$-th aggregated frame being calculated. 
Aggregated FAU features $\varphi_{FAU} \in \mathbb{R}^m$ and head-motion features $\varphi_{head} \in \mathbb{R}^n$ are extracted by the feature extractor $\Phi$:
$
    (\varphi_{FAU}, \varphi_{head})_T = \Phi(\{X_{t_0},...,X_{t}\})
$.
A window of consecutive aggregated frame features is used as input for a data sample, which is labeled with the reward category (i.e., $-5$, $-1$, or $+6$) received during the time step containing the $T$-th aggregated frame. 
The window of aggregated frames begins at the ($T$-$k$)-th and ends at the ($T$+$\ell$)-th aggregated frame. Since some reactions happen after an event, future data is needed to make a prediction for the current event; hence the prediction has a fixed time delay defined by the window.
FAU features and the head-motion features are encoded separately: the temporal series for each is flattened into a single vector and then encoded with a linear layer. 
The two encodings are then concatenated into a single vector, which is input to a multilayer perceptron (MLP).
We include an auxiliary task of predicting the corresponding annotations $\{A_{(T-k)},...,A_{(T+\ell)}\}$ as a single flattened vector $\bm{a} \in \{0,1\} ^{10(k+\ell+1)}$, in which each binary element of $A$ indicates whether a reaction gesture is occurring.
This auxiliary task is intended to speed representation learning and act as a regularizer. 
Empirically, use of this auxiliary task achieves the best test loss but is unnecessary for better-than-random performance in the reward-ranking task (see Section~\ref{sec:results}). We also use a binary classification loss that combines the two negative reward classes as one, which reintroduces the ordinality of the reward classes by additionally penalizing predictions with the wrong sign.
Let $g_\theta(\cdot)$ represent the MLP-based network, $\bm{z} \in \mathbb{R}^c$ be the output (unnormalized log probabilities of the $c$ classes with a corresponding ground-truth one-hot label $\bm{y} \in \{0,1\}^c$), and $\bm{o}$ denote the output of the auxiliary task. $\bm{y}_{bin}$ is the ground-truth binary class, and $\bm{z}_{bin}$ denotes the corresponding binary prediction computed from $\bm{z}$. Therefore,
$(\bm{z}, \bm{o})_T = g_\theta( \{(\varphi_{FAU}, \varphi_{head})_{T-k},...,(\varphi_{FAU}, \varphi_{head})_{T+\ell}\})$.

The loss to be optimized is then expressed as:
\begin{align*}
    \mathcal{L}(\theta) = - \bm{y} \cdot \log(\text{softmax}(\bm{z})) - \lambda_1 \bm{y}_{bin} \cdot \log(\text{softmax}(\bm{z}_{bin})) + \lambda_2 || \bm{a} - \bm{o}||_2
\end{align*}
The neural network is trained with Adam~\cite{kingma2014adam}. 
We employ random search~\cite{bergstra2012random} to find the best set of hyper-parameters to use, including the input's window size ($k$ and $\ell$), learning rate, dropout rate, loss coefficients ($\lambda_1$ and $\lambda_2$), and the depth and widths of the MLP. 
The set of candidate window sizes for random search was informed by ad hoc analysis of the annotations of high-level human facial reactions (Appx.\ref{Annotation Visualization}). Since our dataset is small, we employ k-fold cross validation for the random search of hyper-parameters after randomly sampling one episode of data from each subject into a \textit{holdout} set for final evaluation. Each set of randomly sampled parameters is evaluated across train-test data folds, and the set with the lowest average test loss is selected. Details of the random search process and an ablation study of the reaction mapping design can be found in Appx.\ref{Model Design}.

%% file: sections/5.results.tex
\vspace{-0.04cm}
\section{Results and Evaluation}
\label{sec:results}
\vspace{-0.1cm}


To validate that the learned mappings from our instantiation of stage 1 effectively enable task learning in stage 2, we test the following hypotheses (in which we refer to observers from stage 1 who have created data in the training set as ``\textit{known subjects}''): 

\vspace{-0.05cm}
\textbf{Hypothesis 1} [deployment setting is the same as training setting].~The learned reaction mappings will outperform uniformly random reward ranking, using reaction data from \textit{known subjects} watching the \textit{Robotaxi} task. 

\vspace{-0.1cm}
\textbf{Hypothesis 2} [generalizing \textbf{H1} to online data from novel subjects].~The learned reaction mappings will improve the online policy of a \textit{Robotaxi} agent via updates to its belief over reward functions, based on \textit{online} data from \textit{novel} human observers;

\vspace{-0.1cm}
\textbf{Hypothesis 3} [generalizing \textbf{H1} to a different deployment task].~The learned reaction mappings can be adapted to evaluate robotic-sorting-task trajectories and will outperform uniformly random guessing on return-based rankings of these trajectories, using reaction data from \textit{known subjects}.

\begin{figure*} [b]
\vspace{-0.49cm}
\begin{minipage}{0.715\textwidth}
      \centering
    \includegraphics[width=0.94\textwidth]{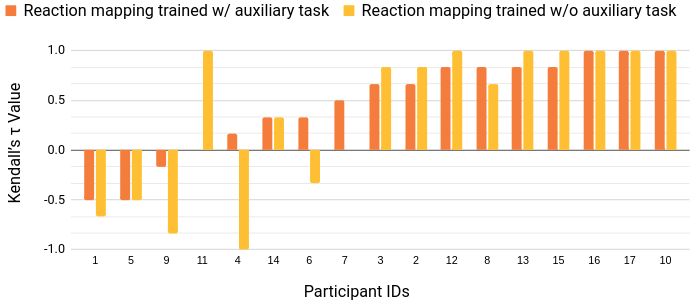}
    \vspace{-0.1cm}
    \caption{Sorted per-subject Kendall's $\tau$ for \textit{Robotaxi} reward-ranking task}
    \label{fig:robotaxi_tau}
\end{minipage}
~
\begin{minipage}{0.27\textwidth}
     \centering
    \includegraphics[width=0.98\textwidth]{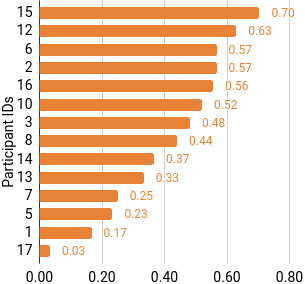}
    \caption{\small Sorted per-subject Kendall's $\tau$ for evaluating robotic sorting trajectories}
    \label{fig:robotic_tau}
\end{minipage}
\end{figure*}

\textbf{Reward-ranking Performance in the Robotaxi Domain}~~
The learned reaction mappings are evaluated on the reward-ranking task. Let $q$ be the random variable for reward event and $x$ be the variable for human reactions. Let $m$ be the discrete random variable over possible reward functions, which in the \textit{Robotaxi} task can be considered a reward ranking. The model $g_\theta(\cdot)$ effectively models $\text{P}(q~|~x, m)$, which is the probability of an event given the human's reaction and a fixed reward ranking $m$. The goal is to find the posterior distribution over $m$:
$\text{P}(m ~|~ q, x) \propto \text{P}(q ~|~ x, m)\text{P}(m)$ (see proof in Appx.\ref{rewardrankingcomputation}).
Given a uniform prior over $m$, we can find $\text{P}(m ~|~ q, x)$ using prediction of the mapping $g_\theta(\cdot)$. The maximum a posteriori reward ranking is chosen as the learned mapping's single estimation after incorporating mappings from all human reaction data in an episode.
To reduce the effect of stochasticity in training neural networks, we train 4 times and report the mean performance.
Fig.~\ref{fig:robotaxi_tau} shows the learned reaction mappings' (per-subject) performance on the episodes in the \textit{holdout} test. 
Using Wilcoxon Signed-Rank test, the mappings' performance on the \textit{holdout} set is significantly better than uniformly random guessing ($\tau = 0$), supporting \textbf{H1}; $p= 0.0024$ with the annotation-reliant auxiliary task and $p=0.0207$ without it. 


\textbf{Online Learning in the Robotaxi Domain}~~The learned reaction mapping can interactively improve an agent's policy: the agent updates its belief over all possible reward rankings using human reactions to its recent behavior and then follows a policy that is approximately optimal with respect to the most likely reward function.
To test such online policy learning, all data collected in stage 1 trains a single reaction mapping, and this reaction mapping is used for single-episode sessions with human observers, none of whom created data within the stage-1 training set. 9 of the 10 participants' interactions achieved a better return than that of a random policy, and 7 of the 10 participants' interactions ended with the probability of reward mappings that lead to optimal behaviors being the highest, moderately supporting \textbf{H2}. Details of this preliminary evaluation can be found in Appx.\ref{Online Learning Results}.

\textbf{Trajectory Ranking in Robotic Sorting Domain}~~
To generalize the reaction mapping trained in the \textit{Robotaxi} domain to the robotic sorting task, we modify the original loss function by removing the multi-class reward classification loss and interpret the reaction mapping's binary output as a ``positivity score'' for an aggregated frame. 
Each human participant observed 7 trajectories (an episode each), chosen from 8 distinct predetermined trajectories. 
Each trajectory accrues return of $+2$ (recycling a can); $-1$ (recycling any other object); or $0$ (nothing placed in the bin). This return enables ground-truth rankings of trajectories. Because we suspect humans react to higher-level actions in this task---to \textit{pick and place object X} rather than to the joint torques applied at 25 ms time steps---the window size of the \textit{Robotaxi} reaction mapping is too small to contain all relevant facial reactions. To address this apparent temporal incompatibility, we compute a per-trajectory positivity score as the mean of the positivity scores of its aggregate frames. A derivation of this approach is in Appx.\ref{RoboticEval} with further details of the trajectory design.
Fig.~\ref{fig:robotic_tau} shows Kendall's $\tau$ values for per-participant rankings of trajectories.
For each trajectory, we compute an overall (cross-subject) positivity score as the mean of the trajectory's per-subject positivity scores. After ranking the 8 trajectories by these scores, Kendall's $\tau$ independence test yields $\tau=0.70$ ($p=0.034$); this test implicitly compares to uniformly random guessing, since its $\tau=0$.
This result supports \textbf{H3}.


%% file: sections/6.conclusion.tex
\vspace{-0.1cm}
\section{Discussion and Conclusion} 
\label{sec:discussion}
\vspace{-0.1cm}

In this paper we introduce the \textsc{lihf} problem and the \textsc{empathic} framework for \textsc{lihf}. We demonstrate that our instantiation interprets human facial reactions in both the training task and the deployment task. We now discuss the limitations of this work and directions for future investigation.

\textbf{Experimental Design}~
We validate our instantiation of \textsc{empathic} with a single training task and similar testing tasks. An important future extension is to generalize this method to tasks with varying temporal characteristics and reward structures. 
In our current setup, agent actions do not have large long-term consequences on the expected return, however changes in human expectations could significantly affect their reactions. One way to incorporate such information into our current modeling approach is to craft corresponding task environments to explore the use of human facial reactions in predicting the long-term returns of agent behaviors.

\textbf{Human Models}~
Data collected in this work allow us to study reactions of human observers who fix their attention on the agent, whereas in real-world settings human observers are often attending to their own tasks. A natural next step is to extend our experiment setup to a more general scenario, in which we also need to infer the relevance of human reactions to the agent's behavior. Additionally, our instantiation assumes that human reactions were influenced by recent and anticipated agent experience but not by other likely factors, such as changing expectations of agent behavior; explicitly modeling such latent human state may further improve \textsc{lihf}.

\textbf{Data Modalities}~
This work maps from facial reactions to discrete rewards. In future work, other forms of human implicit feedback, such as gaze and gestures, could be included to get a more accurate mapping to different task statistics and better performance in a variety of real-world tasks. 

The above limitations notwithstanding, this paper takes a significant step towards the goal of enabling an agent to learn a task from implicit human feedback. It does so by successful application of a learned mapping from human facial reactions to reward types for online agent learning and for evaluating trajectories from a different domain.

%% file: appendix.tex

\section{Problem Formulation} \label{Formulation}

\begin{figure}[h]
    \centering
       \includegraphics[width=0.95\textwidth]{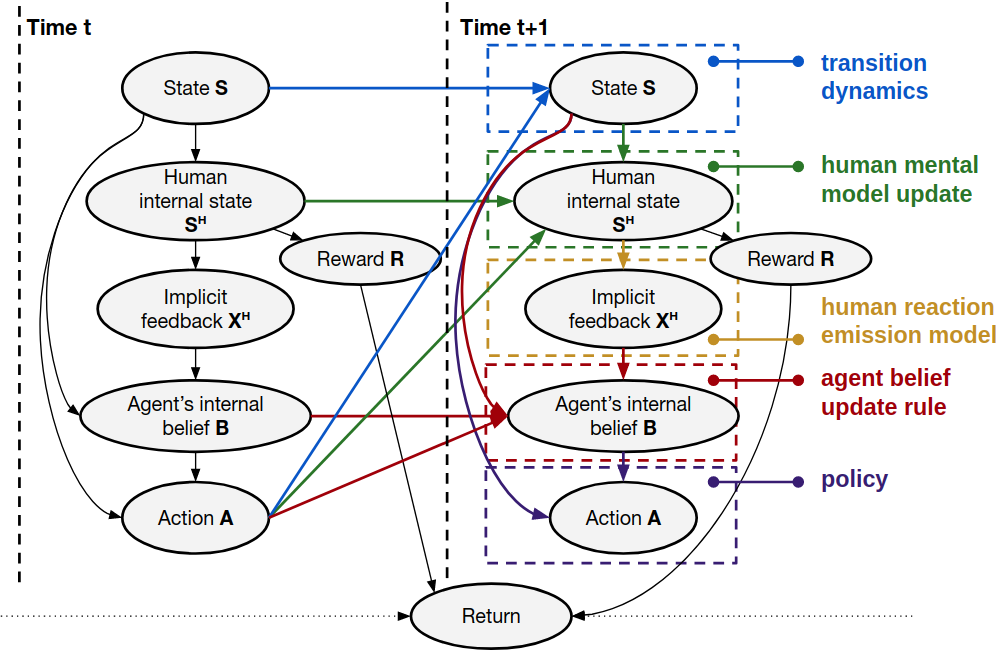}
     \caption{Graphical model for \textsc{lihf} (colored boxes and their identically colored labels represent conditional probability tables)}
     \label{fig:dbn_lihf}
\end{figure}


The graphical model for \textsc{lihf} is shown in Figure~\ref{fig:dbn_lihf}. We assume the human $\mathcal{H}$'s reward function $R^\mathcal{H}$ 
is a temporally invariant element of the human's internal state $S^{\mathcal{H}}$. However, the observation $X^{\mathcal{H}}$ containing implicit human feedback to an action or a trajectory can change over time since it is influenced by the human's mental model of the task and the agent's policy at a particular time step. 
Given an observation $x \in X^{\mathcal{H}}$, the current state $s \in S$, and the previous action $a \in A$, the agent constructs its belief $b \in B$ as a probabilistic memory of arbitrary form and scope over the task domain.
A belief could include, for example, the probability distribution over possible reward functions, which the agent could use to generate a policy (and therefore an action given the current state) that maximizes expected return (aggregated single-step rewards $r \in R$) under the unobserved human reward function $R^\mathcal{H}$. 
Note that reward is not directly dependent on state and action---it is determined by the human entirely (who can, for generality, internally maintain a history of states and actions, and therefore can give non-Markovian rewards). 

\section{Experimental Domains and Data Collection Details} \label{Data Collection Details}

\subsection{Robotaxi}

\textbf{Agent Transition Dynamics}~~In the 8$\times$8 grid-based map, the agent has three actions available at each timestep: maintain direction, turn left, or turn right. When the agent runs into the boundary of the map, it is forced to turn left or right, in the direction of the farther boundary. 

\textbf{Rewards}~~There are three different types of objects associated with non-zero rewards when encountered in the \textit{Robotaxi} environment: if the agent picks up a passenger, it gains a large reward of $+6$; if it runs into a roadblock, it receives a small penalizing reward of $-1$; if the agent crashes into a parked car, it receives a large penalizing reward of $-5$. All other actions result in $0$ reward.

\textbf{Object Regeneration}~~At most 2 instances of the same object type are present in the environment at any given time. An object disappears after the agent moves onto it (a ``pickup''), and another object of the same type is spawned at a random unoccupied location 2 time steps after the corresponding pickup. 

\textbf{Agent Policy}~~The agent executes a stochastic policy by choosing from a set of 3 pseudo-optimal policies under 3 different reward mappings from objects to the 3 reward values:
\begin{itemize}[leftmargin=2.5cm]
    \vspace{-0.1cm}
    \item Go for passenger: ~\{\text{passenger:} $+6$, \text{road-block:} $-1$, \text{parked-car:} $-5$\}
    \item Go for road-block: \{\text{passenger:} $-1$, \text{road-block:} $+6$, \text{parked-car:} $-5$\}
    \item Go for parked-car: \{\text{passenger:} $-1$, \text{road-block:} $-5$, \text{parked-car:} $+6$\}
    \vspace{-0.1cm}
\end{itemize}
The pseudo-optimal policies are computed in real time via value iteration (discount factor $\gamma=0.95$) on a \textit{static} version of the current map, meaning that objects neither disappear nor respawn when the agent moves onto them. We simplify the state space in this manner because the true state space is too large to evaluate and would create too large of a Q function to store, yet this simplification finds an near-optimal policy that almost always takes the shortest path to an object of the type that its corresponding reward function considers to be of highest reward. At the start of an episode and after each pickup, the agent selects 1 of these 3 policies. The agent follows the selected policy until the next pickup, except that there is a $0.1$ probability at each time step that the agent will reselect from the 3 policies. This $0.1$ probability of the agent changing its plans, in a rough sense, was included because we speculated that it would help increase human reactions by making the agent typically exhibit plan-based behavior but sometimes change course, violating human expectations. All selections among the 3 policies are done uniformly randomly.

\begin{figure}[h!]
\centering
\begin{subfigure}{.46\textwidth}
  \centering
        \includegraphics[height=1.67in]{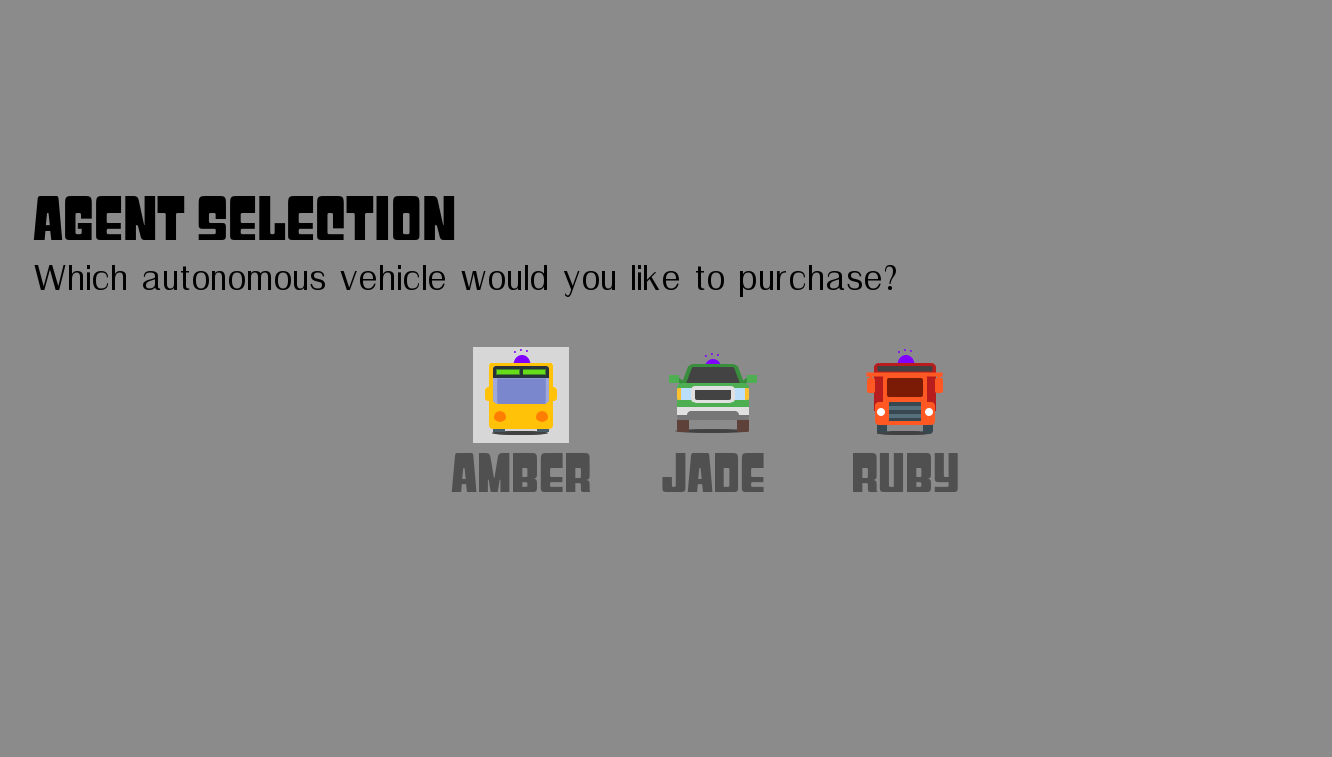}
        \vspace{-0.5cm}
        \caption{Car choice screen}
         \label{fig:carchoice}
\end{subfigure}
~
\begin{subfigure}{.485\textwidth}
   \centering
    \includegraphics[width=\textwidth]{images/robotaxi_env.png}
    \vspace{-0.5cm}
    \caption{\textit{Robotaxi} game view}
    \label{fig:robotaxi_env_appendix}
\end{subfigure}

\caption{\textit{Robotaxi} environment}
\label{fig:fig}
\end{figure}

\subsection{Robotic Sorting Task}

\begin{figure}[h]
    \centering
    \includegraphics[width=0.6\textwidth]{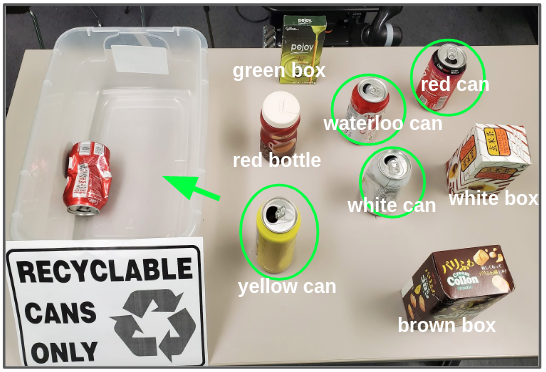}
    \caption{Robotic task table layout with object labels, from the perspective of the human observer}
    \label{fig:robotic_task_layout}
\end{figure}

\begin{figure}[h]
    \centering
    \includegraphics[width=0.98\textwidth]{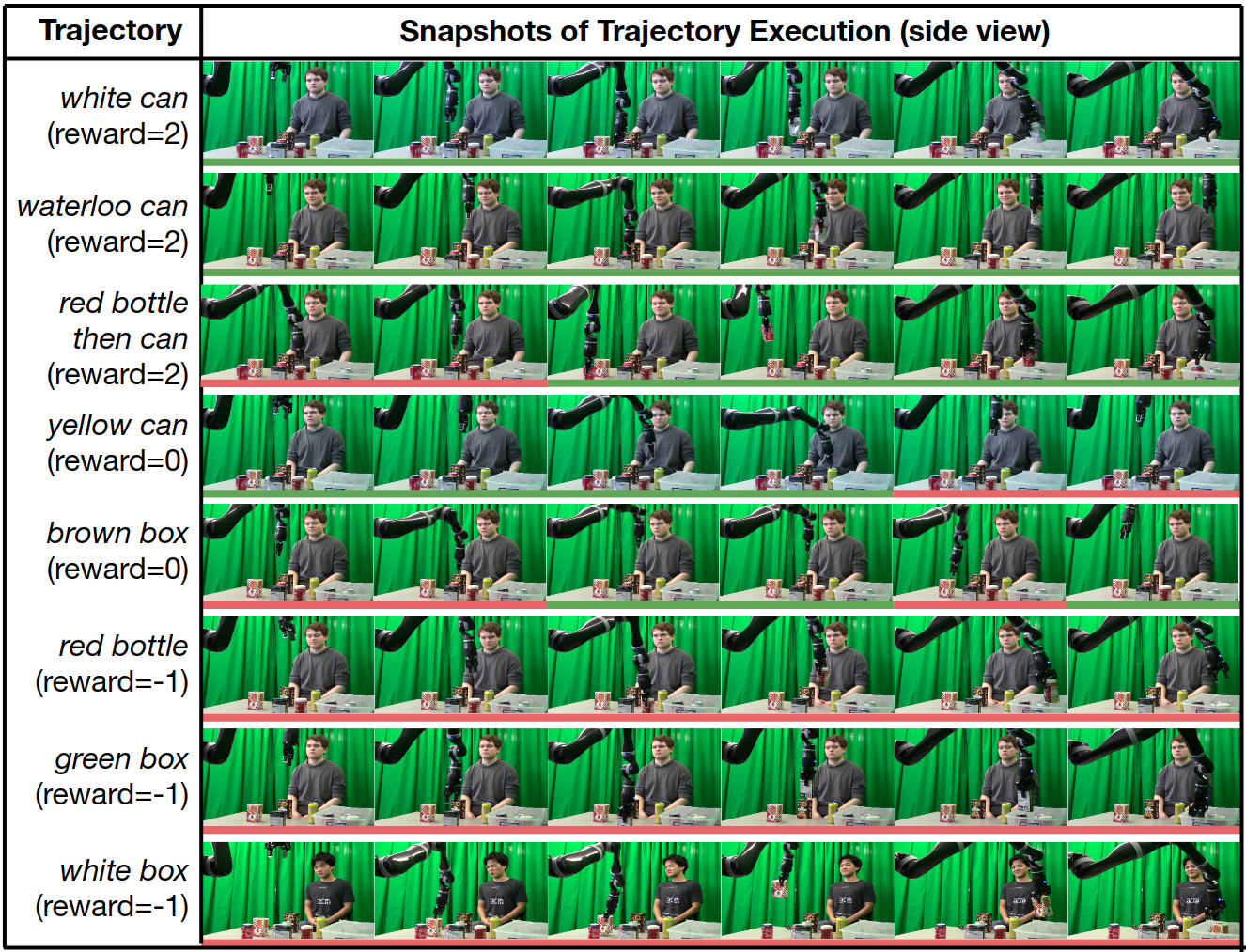}
    \caption{Robotic sorting task trajectories with optimality segmentation}
    \label{fig:robotic_trajectories}
\end{figure}



In the robotic sorting task, the robot executes trajectories programmed through key-frame based kinesthetic teaching. The 7-DOF robotic arm is controlled at 40Hz with torque commands. Fig.~\ref{fig:robotic_task_layout} shows the table layout at the beginning of the task: the robot's task is to sort the aluminum cans into the recycling bin; objects in the green circles give $+2$ rewards when moved into the bin and others give $-1$ rewards.
Fig.~\ref{fig:robotic_trajectories} shows snapshots of the set of 8 arm trajectories we used in the robotic sorting task; each arises from a fixed sequence of torque commands. These fixed torque command sequences produce small variations in the actual trajectory, and any qualitative departure---like an object not being grasped successfully---results in that trajectory being removed from our dataset. 
The 8 episodes each involves 1 or 2 target objects, and ends with a reward of $-1$, $0$ or $+2$. 
Each trajectory of an episode can be further segmented into reaching, grasping, transferring, placing and retracting sub-trajectories. The relative optimality of these sub-trajectories can be determined by whether the projected outcome is desired. For example reaching for a correct object and retracting from picking up a wrong object are both considered optimal while reaching for and transferring a wrong object are both sub-optimal. The optimality of sub-trajectories is also annotated in Fig.~\ref{fig:robotic_trajectories} under each trajectory. Note that our algorithm does not use any such segmentations, which are only for illustration. 

\subsection{Experimental Design}

The instructions we give the participants in \textit{Robotaxi} are as follows:
\begin{itemize}
    \item[--] Hello human! Welcome to Robo Valley, an experimental city where humans don’t work but make money through hiring robots!
    
    \item[--] You’ll start with \$$12$ for hiring \textit{Robotaxi}, and after each session you will be paid or fined according to the performance of the autonomous vehicle or robot.

    \item[--] Your initial \$$12$ will be given to you in poker chips. After each session, we will add or take away your chips based on your vehicle’s score. At the end, you can exchange your final count of poker chips for an Amazon gift card of the same dollar value.
    
    \item[--] For the 3 sessions with a \textit{Robotaxi}, you begin by choosing an autonomous vehicle to lease.
    
    \item[--] The cost to lease one of these vehicles will be \$$4$ each session.
    
    \item[--] The vehicle earns \$$6$ for every passenger it picks up, but it will be fined \$$1$ each time it hits a roadblock  and fined \$$5$ each time it hits a parked car.
    
    \item[--] You will watch the \textit{Robotaxi} earn money for you, and your reactions to its performance will be recorded for research purposes.
    
    \item[--] You will have a chance to practice driving in this world, but the amount earned during the test session won’t count towards your final payout.
    \end{itemize}

The instructions we give the participants in robotic sorting task are as follows:

\begin{itemize}
     \item[--] For the robotic task, the robot is trying to sort recyclable cans out of a set of objects. 

    \item[--] You will earn \$$2$ for each correct item it sorts and get penalized for \$$1$ for each wrong item it puts in the trash bin.

    \item[--] You will watch the robot earn money for you, and your reactions to its performance will be recorded for research purposes.
\end{itemize}

The participants first control the agent themselves for a test session to make themselves familiar with the \textit{Robotaxi} task, removing a source of human reactions changing in ways we cannot easily model.
For the agent-controlled sessions, the participants select an agent at the beginning of each episode of \textit{Robotaxi}. Fig.~\ref{fig:carchoice} shows the view of this agent selection. Unbeknownst to the subject, their selection of a vehicle only affects its appearance, not its policy. This vehicle choice was included in the experimental design as a speculatively justified tactic to increase the subject's emotional investment in the agent's success, thereby better aligning $R$ and $R^\mathcal{H}$ as well as increasing their reactions. At the start of the session, participants are given \$$12$, which they must soon spend to purchase a \textit{Robotaxi} agent before it begins its task. To make their earnings and losses more tangible (and therefore, we speculate, elicit greater reactions), participants are given poker chips equal to their current total earnings. After each session they are paid or fined according to the performance of the agent: their chips are increased or decreased based on the score of \textit{Robotaxi}. At the end of the entire experimental session, participants exchange their final count of poker chips for an Amazon gift card of the same dollar value. 

\subsection{Participant Recruitment}

The participants we recruited are mostly college students in the computer science department. Each participant filled out an exit survey of their backgrounds after completing all episodes of observing an agent. The statistics of these 17 subjects are given below:
\begin{itemize}
    \item Gender: 10 participants are male and 7 are female.
    \item Age: The participants' average age is 20. Ages range from 18 to 28 (inclusive). 
    \item Robotics/AI background: 1 participant is not familiar with AI/robotics technologies at all. 2 have neither worked in AI nor studied it technically, but are familiar with AI and robotics. 13 have not worked in AI but have taken classes related to AI or otherwise studied it technically. Only 1 has worked or done major research in robotics and/or AI.
    \item Ownership of robotics/AI-related products: 7 participants own robotics or AI-related products, while 10 do not. The products include Google Home, Roomba, and Amazon Echo.
\end{itemize}

\section{Annotations of Human Reactions} \label{Annotation Visualization}

\begin{figure}
    \centering
    \centering
    \includegraphics[width=0.9\textwidth]{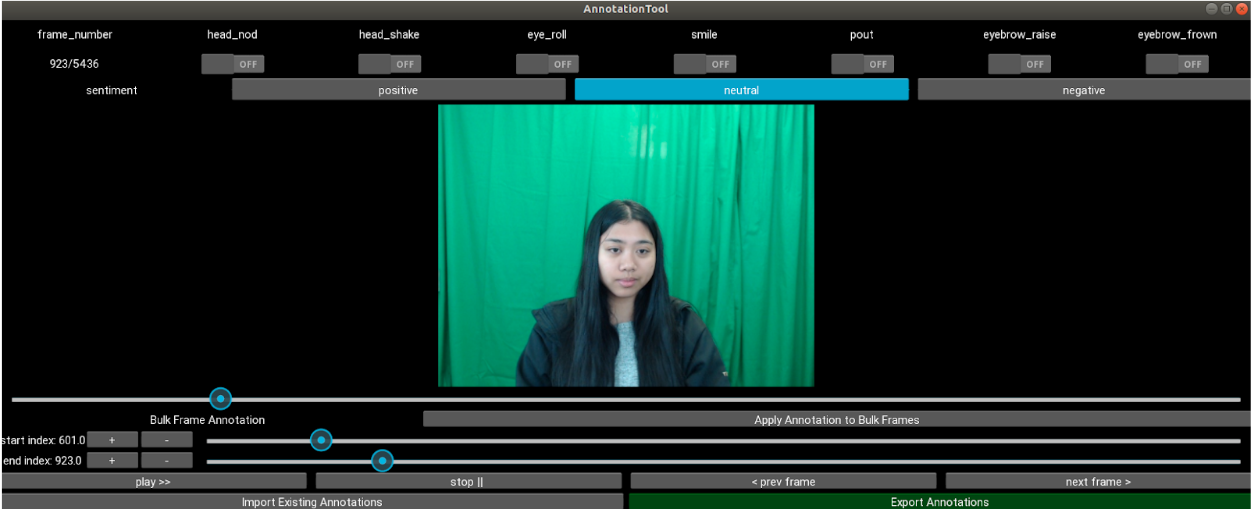}
    \caption{View of the annotation interface. The corresponding trajectory of \textit{Robotaxi} is not displayed.}
    \label{fig:annot_tool}
\vspace{-0.3cm}
\end{figure}

\begin{figure*}
    \centering
      \vspace{0.2cm}
      \centering
        \includegraphics[width=0.4\textwidth]{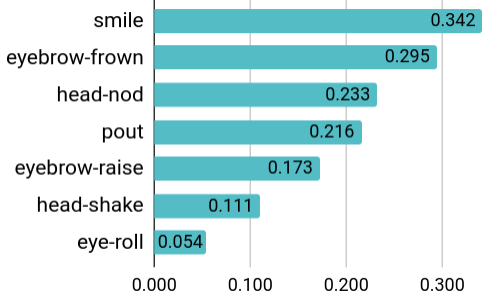}
        \caption{Proportion of annotated gestures}
         \label{fig:gesture_freq}
\end{figure*}
\begin{figure*}
    \centering
    \includegraphics[width=\textwidth]{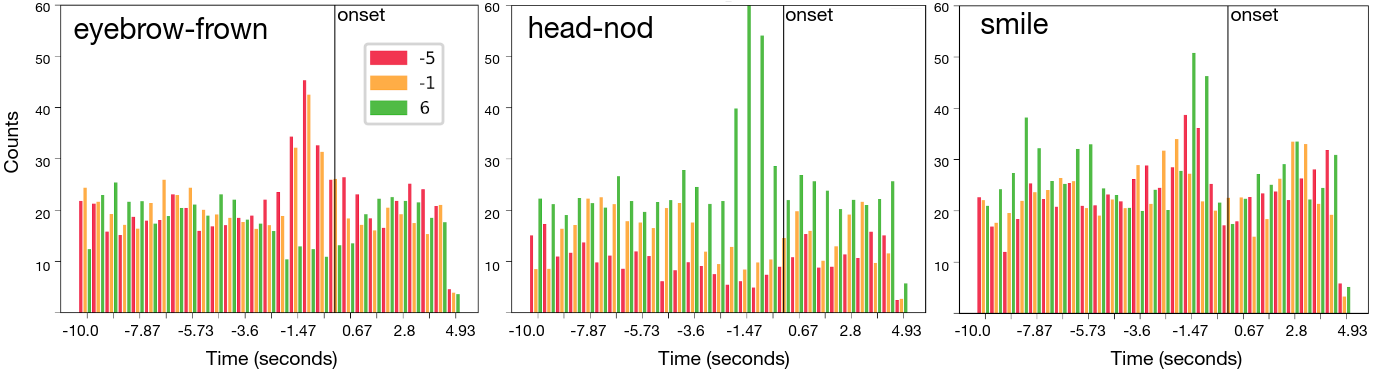}
    \includegraphics[width=\textwidth]{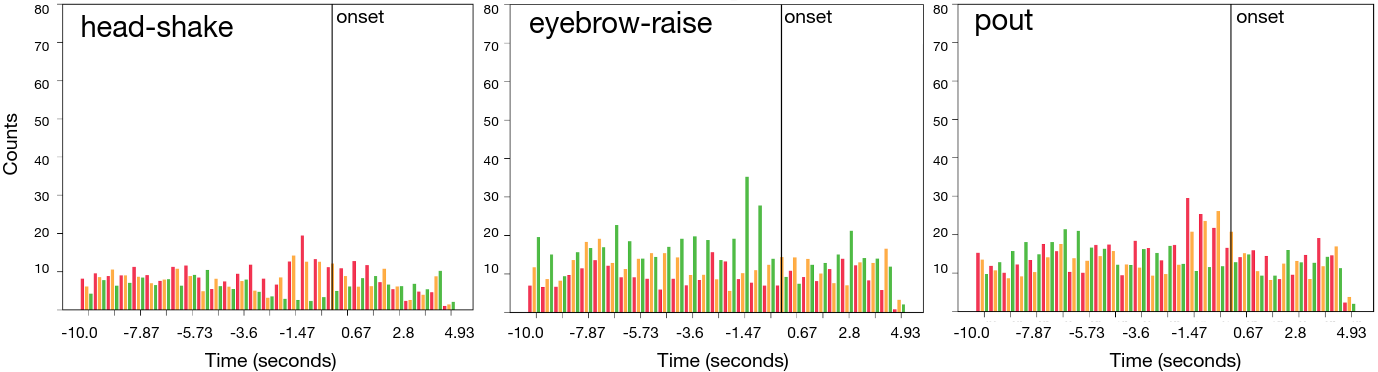}
    \includegraphics[width=\textwidth]{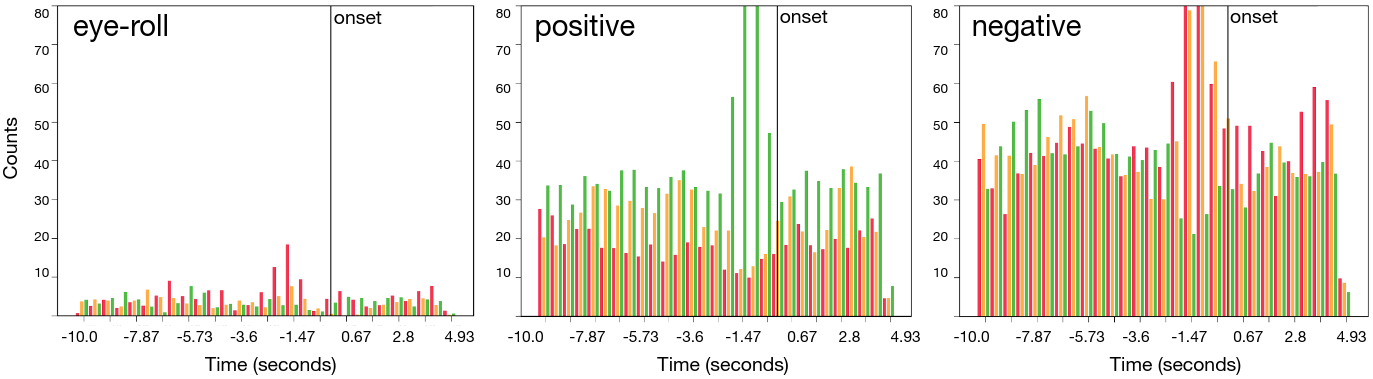}
    \caption{Histograms of non-zero reward events around feature onset}
    \label{fig:feat_hist}
\vspace{-0.3cm}
\end{figure*}


\subsection{The Annotation Interface}


To gain a better understanding of the dataset and the \textsc{lihf} problem as a whole, two of the authors annotated the collected dataset. These annotations are not intended to serve as ground truth and are only used as labels for an auxiliary task in our training of reaction mappings. Therefore, training/calibration of annotators, evaluation of annotators via inter-rator reliability scores, etc. are not important.
The interface for annotating the data is displayed in Fig.~\ref{fig:annot_tool}. A human annotator marks whether facial gestures and head gestures are occurring in each frame, effectively marking the onset and offset of such gestures. Annotation is performed without any visibility of the corresponding game trajectory.
The proportion of 7 reaction gestures in the annotation is shown in Fig.~\ref{fig:gesture_freq}. Annotations provide several benefits in this study: in our search for a modeling strategy, we found our first successful reaction mapping while using annotations directly as the only supervisory labels; annotations provide labels for an auxiliary task to regularize training of and speed representation learning by (both important for a small dataset) the reaction mapping from the features extracted automatically via OpenFace \cite{baltrusaitis2018openface,baltruvsaitis2015cross,zadeh2017convolutional}; and an annotation-based analysis of our data helped us find a temporal window of reaction data around an event that is effective for inferring the reward types for that event.
\vspace{-0.25cm}

\subsection{Visualizations of Annotated Data}
The annotations can be used to visualize the temporal relationship between reaction onsets and events (rewards). Fig.~\ref{fig:feat_hist} shows example histograms of reward events binned into time windows around feature onsets. As demonstrated by Fig.~\ref{fig:feat_hist}, the onsets of certain gestures such as eyebrow-frown and head-nod correlate with negative or positive events respectively (peaking around 1.47s before the onset). While smile accounts for a large portion of overall gestures (Fig.~\ref{fig:gesture_freq}), it does not correlate strongly with either positive or negative events, contradicting the assumption made by several prior studies that smile should always be treated as positive feedback \cite{li2020facial,arakawa2018dqn,zadok2019affect,veeriah2016face}. While this observation could be specific to our experimental setting or domain, it agrees with established research on the emotional meanings of smiles as shown in the work of \citet{hoque2012exploring} and \citet{niedenthal2010simulation}.

In these histograms, the contours of red and yellow bars are strikingly similar in most subplots of Fig.~\ref{fig:feat_hist}, which suggests that although an individual may react differently to the events that provide $-1$ and $-5$ reward, it may be hard to distinguish between them through single gestures. We also find that reactions (across all gestures) are likely to occur between 2.8s before and 3.6s after an event (shown as a peak in the histograms), which we use as a prior for designing the set of candidate time windows that random hyperparameter search draws from (see Appx.\ref{Hyperparameters}).

\section{Feature Extraction}\label{Feature Extraction}
\vspace{-0.3cm}
The specific output data we use from \href{https://github.com/TadasBaltrusaitis/OpenFace/wiki/Output-Format}{OpenFace} \cite{baltrusaitis2018openface, baltruvsaitis2015cross,zadeh2017convolutional} are: 
    [success, AU01\_c ,  AU02\_c ,  AU04\_c ,  AU05\_c ,  AU06\_c ,  AU07\_c ,  AU09\_c ,  AU10\_c ,  AU12\_c ,  AU14\_c ,  AU15\_c ,  AU17\_c ,  AU20\_c ,  AU23\_c ,  AU25\_c ,  AU26\_c ,  AU28\_c ,  AU45\_c ,   AU01\_r ,  AU02\_r ,  AU04\_r ,  AU05\_r ,  AU06\_r ,  AU07\_r ,  AU09\_r ,  AU10\_r ,  AU12\_r ,  AU14\_r ,  AU15\_r ,  AU17\_r ,  AU20\_r ,  AU23\_r ,  AU25\_r ,  AU26\_r ,  AU45\_r ,  pose\_Tx ,  pose\_Ty ,  pose\_Tz ,  pose\_Rx ,  pose\_Ry ,  pose\_Rz ]. 

The AUx\_c signals are outputs from classifiers of activation of facial action units (FAU) and AUx\_r are from regression model that are designed to capture the intensity of the activation of facial action units. The pose\_T and pose\_R signals are detected head translation and rotation with respect to the camera pose. Since the camera pose and relative position of a person with the camera varies from training time to application time, we explicitly model the change in the detected person's head pose by maintaining a running average and subtract the average from all incoming pose features. We then use a time window of the past 50 feature frames and compute the Fourier transform coefficients as the head-motion features we feed into the neural network.

\section{Model Design} \label{Model Design}

\subsection{Data Split of k-fold Cross Validation for Random Search} \label{Data Split}
\begin{figure}[t]
    \centering
    \includegraphics[width=.6\textwidth]{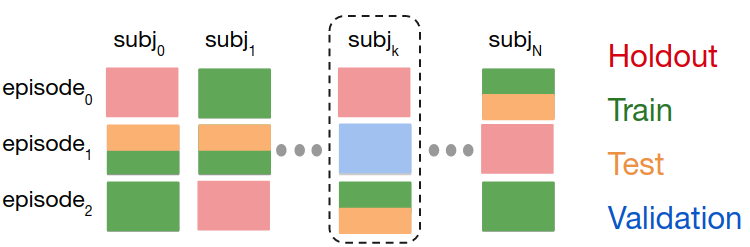}
    \caption{Diagram of data split for subject k}
    \vspace{-0.5cm}
    \label{fig:data_split}
\end{figure}

During our search for hyperparameters that learn an effective reaction mapping from data gathered in \textit{Robotaxi}, we used a data-splitting method designed to avoid overfitting and to have relatively large training sets despite our small dataset size. Recall that each participant observe and react to 3 episodes. Of these 3 episodes, 1 is randomly chosen as a holdout episode and is not used for training or testing except for final evaluation. With the remaining 2 episodes per subject, we split data such that different train-test-validation sets are created for each subject, as shown in Fig.~\ref{fig:data_split}. Specifically, we construct a train-test-validation set for each subject by assigning one episode of the target subject as the validation set, randomly sampling half (either the first half or the second) of an unused episode from each subject into the test set, and using the remaining data in the training set. For a target subject, a model is trained on the subject's corresponding training set and tested after each epoch on the test set. The epoch with the best cross-validation loss is chosen as the early stopping point, and the model trained at this epoch is then evaluated on the validation set. The performance of a hyperparameter set is defined as the mean of the cross entropy losses across each subject's validation set. The hyperparameter set with the lowest such mean cross entropy loss is selected for evaluation on the holdout set. The data split for evaluation on the holdout set is similar but simpler. From the 2 episodes per subject that are not in the holdout set, half an episode is randomly sampled into the test set and the rest are in the training set. A single model is trained (stopping with the lowest cross-entropy loss on test set) and then evaluated on the holdout set. 


\subsection{Hyperparameters}\label{Hyperparameters}

Random search is used to find the best set of hyper-parameters, including input window size ($k$ and $l$), learning rate, dropout rate, loss coefficients ($\lambda_1$ and $\lambda_2$), depth and widths of the MLP hidden layers. Fig.~\ref{fig:feat_hist} indicates that reactions are likely to onset between 2.8s before and 3.6s after an event. Therefore, we convert the corresponding range of temporal window into the number of aggregated frames before and after a particular prediction point (aggregated frame) and use that as the range to sample the input window. 
Each set of randomly sampled parameters is evaluated on all 17 train-test folds and the set with the lowest average test loss is selected. For each model, the weights with the lowest test loss are saved and evaluated on the validation set.

The best hyper-parameters found through random search are: \{learning\_rate = $0.001$, batch\_size = $8$, $k$ = $0$, $l$ = $12$, dropout\_rate = $0.6314$, $\lambda_1$ = $2$, $\lambda_2$ = $1$\}.
Below is the best model architecture found through random search:
\scriptsize{
\begin{verbatim}
(facial_action_unit_input): Linear(in_features=455, out_features=64, bias=True)
(head_pose_input): Linear(in_features=702, out_features=32, bias=True)
(hidden): ModuleList(
    (0): Linear(in_features=96, out_features=128, bias=True)
    (1): BatchNorm1d(128, eps=1e-05, momentum=0.1)
    (2): LeakyReLU(negative_slope=0.01)
    (3): Dropout(p=0.63, inplace=False)
    (4): Linear(in_features=128, out_features=128, bias=True)
    (5): BatchNorm1d(128, eps=1e-05, momentum=0.1)
    (6): LeakyReLU(negative_slope=0.01)
    (7): Dropout(p=0.63, inplace=False)
    (8): Linear(in_features=128, out_features=64, bias=True)
    (9): BatchNorm1d(64, eps=1e-05, momentum=0.1)
    (10): LeakyReLU(negative_slope=0.01)
    (11): Dropout(p=0.63, inplace=False)
    (12): Linear(in_features=64, out_features=8, bias=True)
    (13): BatchNorm1d(8, eps=1e-05, momentum=0.1)
    (14): LeakyReLU(negative_slope=0.01)
    (15): Dropout(p=0.63, inplace=False))
(out): Linear(in_features=8, out_features=3, bias=True)
(auxiliary_task): Linear(in_features=128, out_features=130, bias=True)
\end{verbatim}
}

\normalsize
\subsection{Ablation Study for Predictive Model Design} \label{Ablation}

\begin{figure}[h!]
\centering
\includegraphics[width=0.9\textwidth]{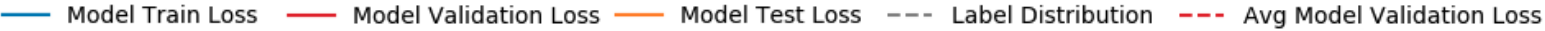}
\begin{subfigure}{.46\textwidth}
  \centering
        \includegraphics[width=\textwidth]{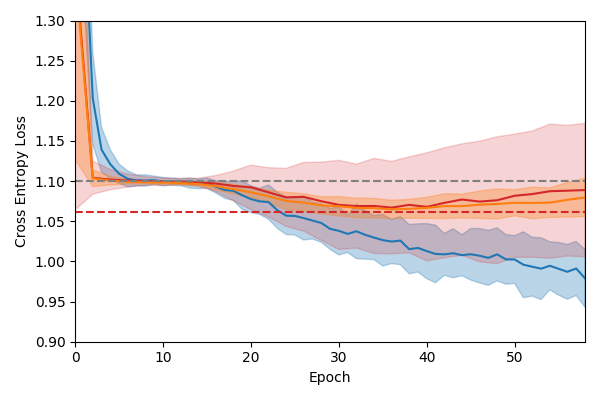}
        \vspace{-0.5cm}
        \caption{Proposed model} 
         \label{fig:model0_loss}
\end{subfigure}
~
\begin{subfigure}{.46\textwidth}
   \centering
    \includegraphics[width=\textwidth]{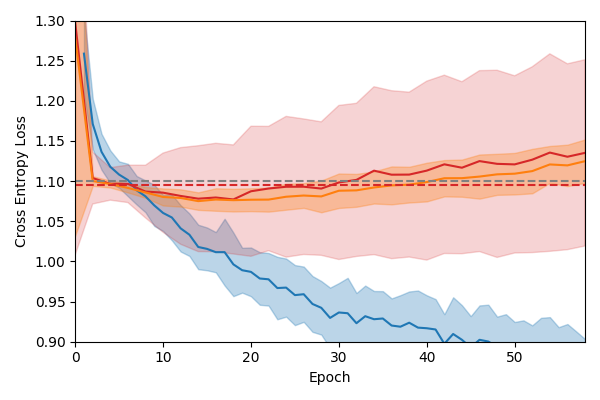}
    \vspace{-0.5cm}
    \caption{Model trained without auxiliary task}
    \label{fig:noaux_loss}
\end{subfigure}
~
\begin{subfigure}{.46\textwidth}
  \centering
        \includegraphics[width=\textwidth]{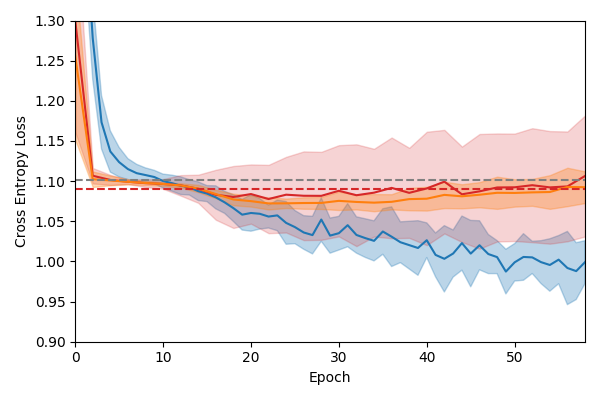}
        \vspace{-0.5cm}
        \caption{Only use FAU features as input} 
         \label{fig:facsonly_loss}
\end{subfigure}
~
\begin{subfigure}{.46\textwidth}
   \centering
    \includegraphics[width=\textwidth]{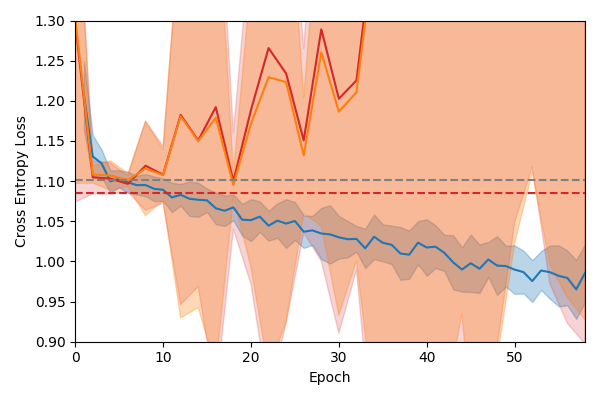}
    \vspace{-0.5cm}
    \caption{Only use head-motion features as input}
    \label{fig:poseonly_loss}
\end{subfigure}
~
\begin{subfigure}{.46\textwidth}
   \centering
    \includegraphics[width=\textwidth]{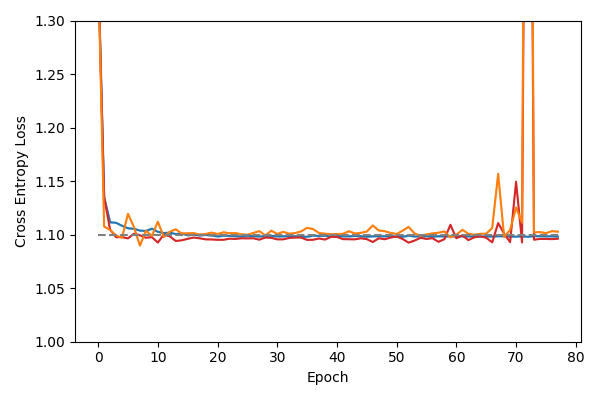}
    \vspace{-0.5cm}
    \caption{End-to-end model (Resnet+LSTM architecture)}
    \label{fig:end2end}
\end{subfigure}
\caption{Loss profiles for training different models (each model has its own set of best hyper parameters found through random search except the end-to-end model)}
\label{fig:ablation_loss}
\vspace{-0.5cm}
\end{figure}

To validate the effectiveness of our model design, we conduct ablation study on the use of auxiliary task and input features. Fig.~\ref{fig:ablation_loss} shows the loss profiles during training across 17 subject train-test-validation sets for the proposed model, the model without auxiliary loss, the model using only FAU features, and the model using only head-motion features respectively. Each of them uses its own set of hyperparameters found through random parameter search. All models are evaluated using 17-fold cross validation based on each subject, and the set with the lowest average test loss is selected. As shown in Fig.~\ref{fig:ablation_loss}, our best full model has the best average loss on the test set, and also has the lowest mean and variance of validation loss compared with the other three models.
We also tested training an end-to-end model with a Resnet-18 CNN as feature extractor and an LSTM model for processing features within a window. The CNN-LSTM model's training loss did not decrease to be lower than that obtained by outputting the label distribution. Given the size of this end-to-end model, we could not efficiently conduct an extensive hyperparameter search and have to rely on manual tuning. We speculate that as a main factor of failure. Meanwhile, we may not have enough data to effectively train a CNN-based feature extractor. Leveraging existing models such as OpenFace \cite{baltrusaitis2018openface, baltruvsaitis2015cross,zadeh2017convolutional} for extracting features alleviates our modeling burden with limited amount of data.

\section{Computing Reward Ranking with Learned Reaction Mapping} \label{rewardrankingcomputation}
$q$ is the random variable for reward event and $x$ is the variable representing human reactions. $m$ is the discrete random variable over possible reward functions (i.e. reward rankings). The learned mapping effectively models $\text{P}(q~|~x, m)$, which is the probability of an event given the human's reaction and a fixed reward ranking $m$. The goal is to find the posterior distribution over $m$: $\text{P}(m ~|~ q, x)$.

Below is the proof for $\text{P}(m ~|~ q, x) \propto \text{P}(q ~|~ x, m)\text{P}(m)$:
\begin{align}
    \text{P}(q, x, m) & = \text{P}(q ~|~ x, m)\text{P}(x~|~m)\text{P}(m) \\
                    & = \text{P}(m ~|~ q, x) \text{P}(q, x) 
\end{align}
\begin{equation}
    \text{P}(m ~|~ q, x) \text{P}(q~|~x)\text{P}(x) = \text{P}(q ~|~ x, m)\text{P}(x~|~m)\text{P}(m) 
\end{equation}
$x$ and $m$ are conditionally independent since the human observes the reward, therefore: 
\begin{equation}
    \text{P}(x~|~m) = \text{P}(x)
\end{equation}
Hence,
\begin{equation}
    \text{P}(m ~|~ q, x) \text{P}(q~|~x) = \text{P}(q ~|~ x, m)\text{P}(m) 
\end{equation}
$\text{P}(q~|~x)$ is constant across all values of $m$, therefore: $\text{P}(m ~|~ q, x) \propto \text{P}(q ~|~ x, m)\text{P}(m)$.

\section{Reaction Mapping Training Profile}
\label{training}

\begin{figure}[h!]
\centering
\includegraphics[width=0.8\textwidth]{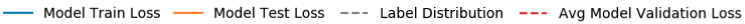}
\begin{subfigure}{.46\textwidth}
  \centering
        \includegraphics[width=\textwidth]{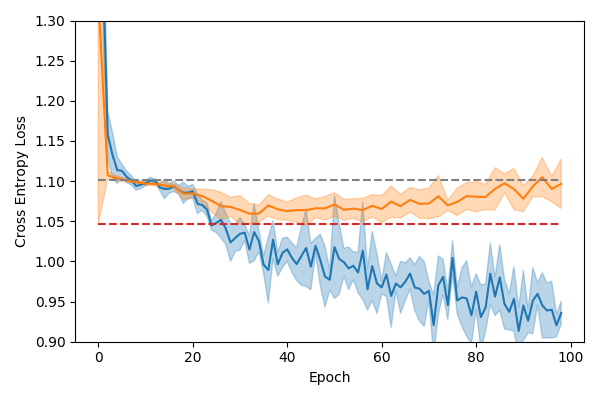}
        \vspace{-0.5cm}
        \caption{Holdout data as evaluation set} 
         \label{fig:singlemodel_holdout}
\end{subfigure}
~
\begin{subfigure}{.46\textwidth}
   \centering
    \includegraphics[width=\textwidth]{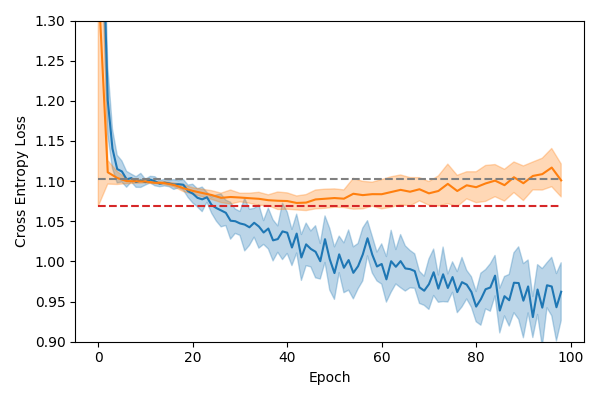}
    \vspace{-0.5cm}
    \caption{Human proxy episodes as evaluation set}
    \label{fig:singlemodel_hpt}
\end{subfigure}
\caption{Loss profiles for training final models for reward ranking evaluation.}
\label{fig:single_model_loss}
\end{figure}
Fig.~\ref{fig:single_model_loss} shows the loss profiles (across 4 repetitions) for training final (single-model) mappings to be evaluated on our stage-2 instantiations. The red dotted line shows the average validation loss across 4 repetitions using the model selected with the lowest testing loss.

\section{Reward Ranking Performance on Human Proxy Test Episodes}
\label{hpt}

\begin{figure}[h!]
      \centering
    \includegraphics[width=0.8\textwidth]{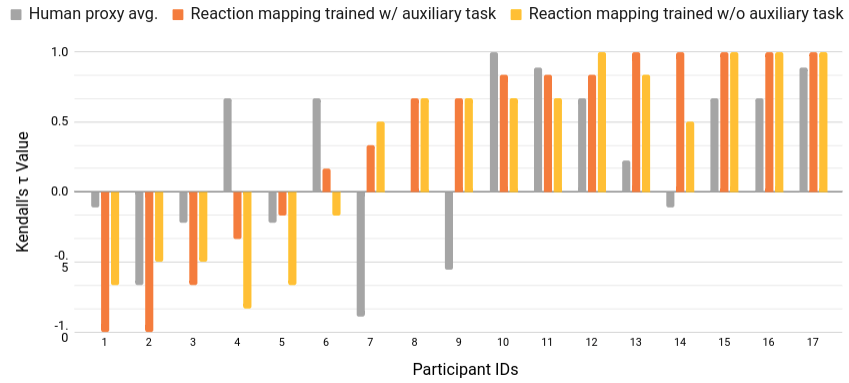}
    \caption{Sorted per-subject Kendall's $\tau$ for \textit{\textit{Robotaxi}} reward-ranking task on the human proxy test episodes}
    \label{fig:robotaxi_tau_holdout}
        \vspace{-0.15cm}
\end{figure}

The result of performing the same reward ranking task on the human proxy test episodes is shown in Fig.~\ref{fig:robotaxi_tau_holdout}. In this setting, the episodes on which the mappings are evaluated are the same as the human proxies viewed, and the rest of the episodes are used as training data. In general, our model's performance is bad on participants that the human proxies also find difficult and good on participants the human proxies performed well on, with a few exceptions.   

\section{Effects of Different Belief Priors}

\begin{figure}[h!]
\centering
\includegraphics[width=0.8\textwidth]{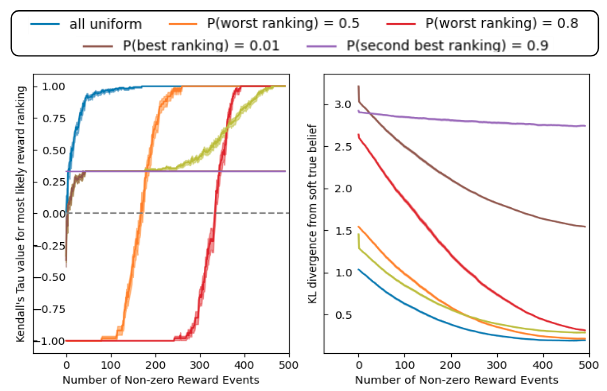}
\caption{Performance of reward inference starting from different priors over reward rankings.}
\label{fig:wrong_priors}
\end{figure}

To test whether our reward learning method can recover from prior beliefs over reward rankings that give low probability to the correct reward ranking
, we perform inference over all possible reward rankings starting from different priors. We pool predictions of the learned \textit{Robotaxi} reaction mapping on the holdout dataset. From this pool of likelihoods, we randomly sample without replacement to update the reward belief. Different classes of reward ranking priors are tested:
\begin{itemize}
    \item \textbf{all uniform}: uniform prior over all possible reward rankings (used in all other experiments); 
    \item \textbf{P(worst ranking) = $p$}: the reward mapping that ranks events in the \textit{reverse} of the correct ranking has prior probability mass $p$, and the rest of the reward rankings uniformly share $1-p$ probability mass; 
    \item \textbf{P(best ranking) = $p$}: the correct reward ranking has prior probability mass $p$, and the rest of the reward rankings uniformly share $1-p$ probability mass; and
    \item \textbf{P(second best ranking) = $p$}: the reward ranking that correctly ranks the positive-reward event first but incorrectly rank the two negative-reward events has prior probability mass $p$, and the rest of the reward rankings uniformly share $1-p$ probability mass.
\end{itemize}
As a function of the number of non-zero reward observations used for inference, we record the following performance metrics: the average Kendall's $\tau$ score of the most likely reward ranking, and the KL divergence between the current belief and a soft true belief distribution, defined such that the prior probability of a particular reward ranking is 
$\exp(\lambda\tau)/Z$, where $\tau$ is the Kendall's $\tau$ value for that reward ranking.
This experiment is repeated 100 times and the mean performance over the number of non-zero reward events is shown in Fig.~\ref{fig:wrong_priors}.

Out of the six different priors we tested, the hardest belief prior to recover from is \textbf{P(second best ranking) = 0.9}, where the two negative rewards are swapped compared to the correct ranking. In four out of the six cases, the reward inference process is able to recover the true reward ranking after incorporating enough data points. If higher weight is put on the predicted likelihood, the inference will converge faster to the correct reward ranking. The reward inference process in general is sensitive to the prior used. Therefore, when lacking justification for a biased prior, we recommend starting with a uniform prior.

Note that this experiment is conducted offline with data from human observers watching an agent with a fixed policy. We expect the learning dynamics to be different when the agent is updating its behavior policy online according to its existing belief, using live data from the human observer.

\section{Online Learning Results} \label{Online Learning Results}

\begin{figure}[h!]
\centering
\includegraphics[width=0.8\textwidth]{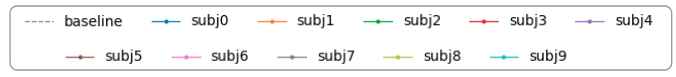}
\vspace{0.2cm}
\begin{subfigure}{.48\textwidth}
  \centering
  \includegraphics[width=\linewidth]{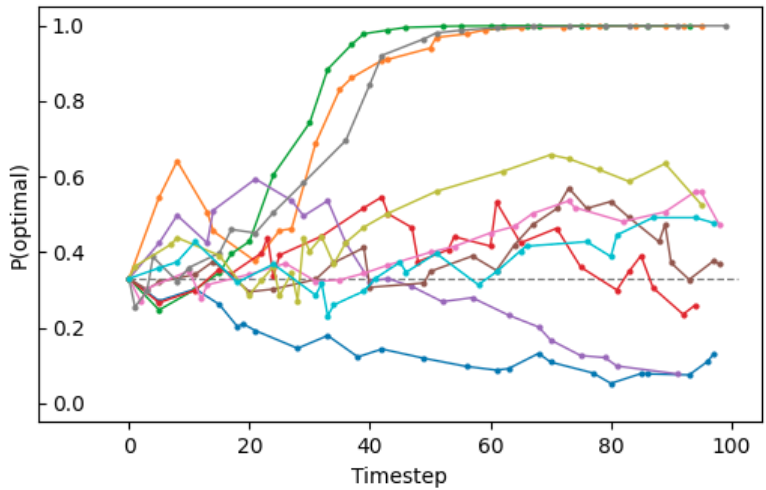}  
  \caption{\small Probability}
  \label{fig:sub-first}
\end{subfigure}
~~
\begin{subfigure}{.48\textwidth}
  \centering
  \includegraphics[width=\linewidth]{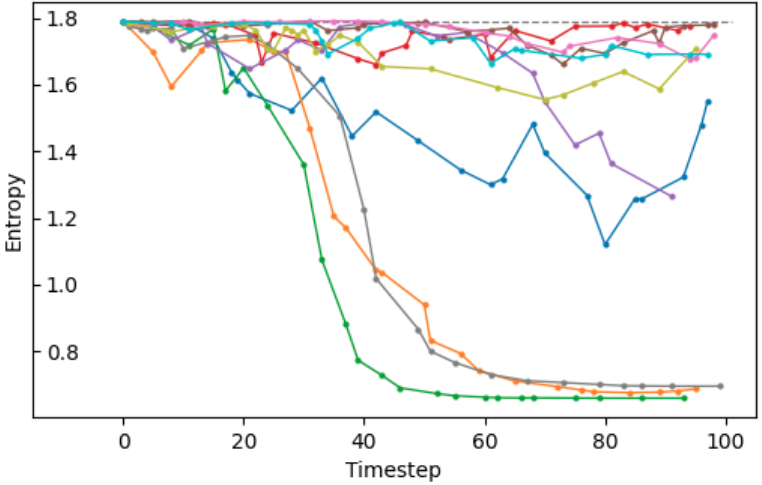}  
  \caption{\small Entropy over rewards}
  \label{fig:sub-second}
\end{subfigure}
\vspace{0.2cm}
\begin{subfigure}{.48\textwidth}
  \centering
  \includegraphics[width=\linewidth]{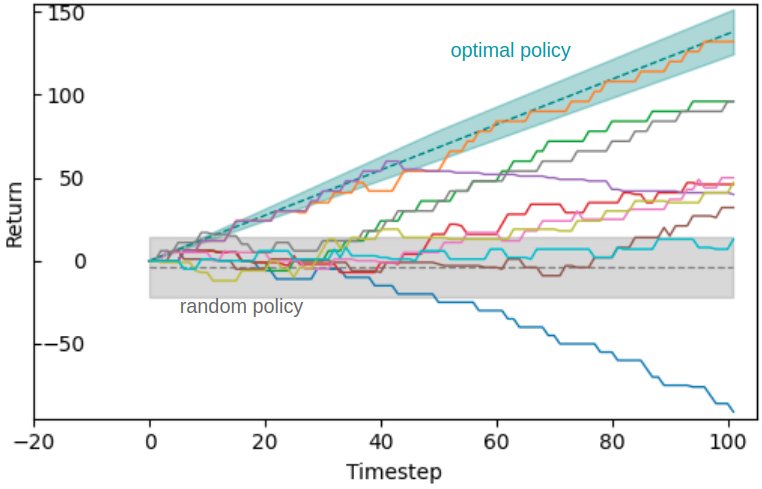}  
  \caption{\small Return}
  \label{fig:sub-second}
\end{subfigure}
~~
\begin{subfigure}{.48\textwidth}
  \centering
  \includegraphics[width=\linewidth]{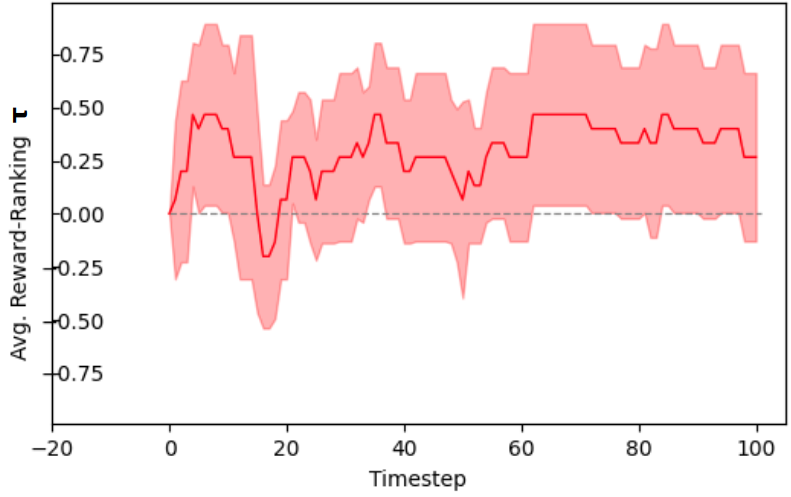}  
  \caption{\small Kendall's $\tau$ values}
  \label{fig:sub-second}
\end{subfigure}
 \caption{Trials of informal online learning sessions in \textit{Robotaxi}}
\label{fig:online_learning_result}
\vspace{-0.3cm}
\end{figure}

This online learning evaluation is conducted in informal settings. Due to the practice of social distancing during the world-wide spread of COVID-19, the authors recruited their friends as test subjects and conducted the evaluation in their own homes. Because of this informality, two aspects of the experimental design were not followed. First, participants were unpaid, removing our main mechanism for aligning $R^\mathcal{H}$ and $R$. (Note: $R$ never produces reward observed by the agent but is instead used only for evaluating agent performance.) Second, prior to human subjects acting as observers, they did not control the agent for an episode, an experimental step that had been intended to provide the subjects an understanding of the task. We suspect that the lack of each of these aspects worsens our results; nonetheless, these results are fairly positive.

As shown in Fig.~\ref{fig:online_learning_result}, in 9 out of 10 trials the final return is positive (8 out of 10 significantly higher than random policy returns) with $p = 0.0134$ by binomial test,
and in 3 out of 10 trials the belief converged to the optimal and the second optimal reward rankings (both ranking passenger pick-up highest) with low entropy (when the probability mass evenly splits between the two rewards, the entropy is $-\ln(0.5)*0.5*2=0.69$). Further, the average Kendall's $\tau$ value of reward ranking is higher than the random guessing baseline, after a small number of initial timesteps. 

\section{Preliminary Modeling of Other Task Statistics} \label{RLstats}

We also attempted to model other task statistics including the following ones computed with the agent's behavior policy $\pi^b$ and the optimal policy $\pi^*$ under the ground-truth reward function:
\begin{itemize}
    \item Q-value of an action under optimal policy:  $ Q^*(s,a)= R(s,a) + V^*(s')$
    \item Optimality (0/1) of an action ($\mathds{1}$ is the indicator function): $ O(s,a) = \mathds{1}_{[Q^*(s,a)]}(Q(s,a)) $
    \item Q-value of an action under the behavior policy:  $ Q^{\pi^b}(s,a)= R(s,a) + V^*(s')$
    \item Advantage of an action under optimal policy: $ A^*(s,a) = Q^*(s,a) - V^*(s)$
    \item Advantage of an action under behavior policy: $ A^{\pi^b}(s,a) = Q^{\pi^b}(s,a) - V^{\pi^b}(s)$
    \item Surprise modeled as the difference in $Q$: $ S(s,a) = Q^{\pi^b}(s,a) - Q^*(s,a) $
\end{itemize}

As mentioned previously, for computing the agent's policy in \textit{Robotaxi}, we use an approximate optimal policy by assuming the grid map is static and run value iteration on the gridworld map (we repeat value iteration computation whenever the map changes, i.e. some object was picked up). We believe this policy is optimal as long as there are no more than 2 objects of the same type in the map. We then use Monte Carlo rollouts to estimate the value and Q-value along each trajectory.

\begin{figure}[ht]
\centering
\includegraphics[width=0.9\textwidth]{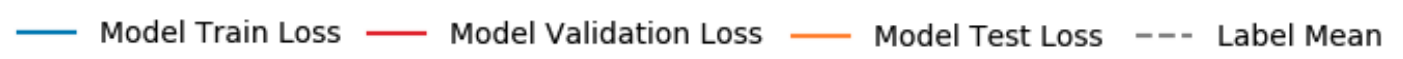}
\begin{subfigure}{.48\textwidth}
  \centering
  \includegraphics[width=\linewidth]{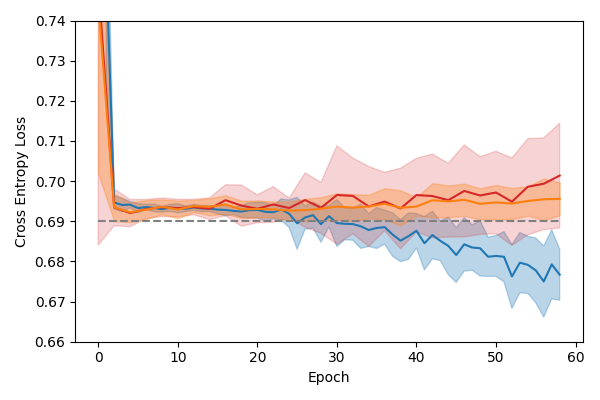}  
  \caption{ $O(s,a)$}
  \label{fig:rlstats2}
\end{subfigure}
\begin{subfigure}{.48\textwidth}
  \centering
  \includegraphics[width=\linewidth]{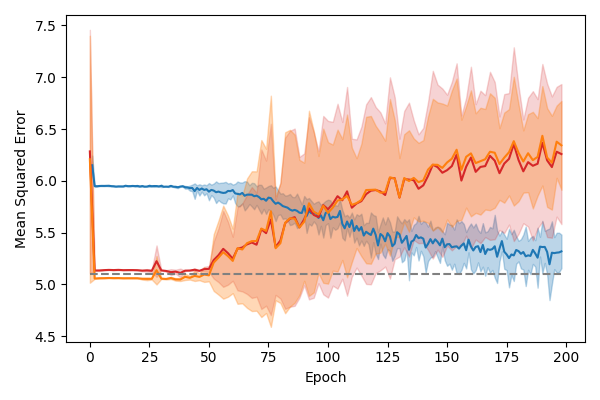}  
  \caption{ $Q^*(s,a)$}
  \label{fig:rlstats3}
\end{subfigure}
\begin{subfigure}{.48\textwidth}
  \centering
  \includegraphics[width=\linewidth]{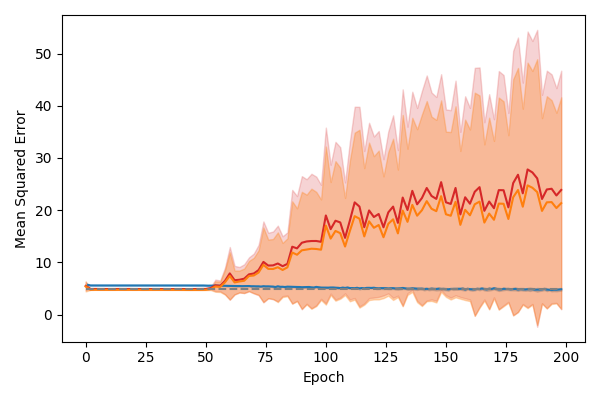}  
  \caption{ $Q^{\pi^b}(s,a)$}
  \label{fig:rlstats4}
\end{subfigure}
\begin{subfigure}{.48\textwidth}
  \centering
  \includegraphics[width=\linewidth]{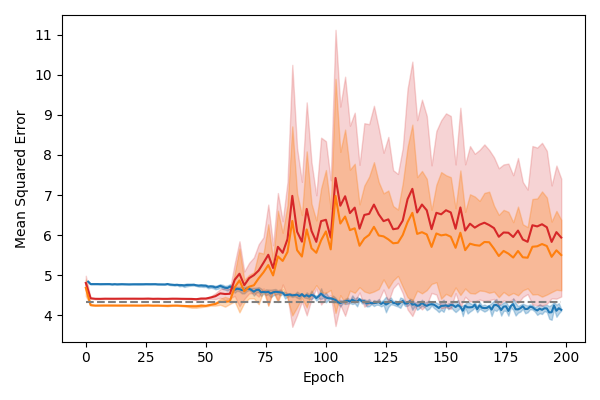}  
  \caption{$ A^*(s,a)$}
  \label{fig:rlstats5}
\end{subfigure}
\begin{subfigure}{.48\textwidth}
  \centering
  \includegraphics[width=\linewidth]{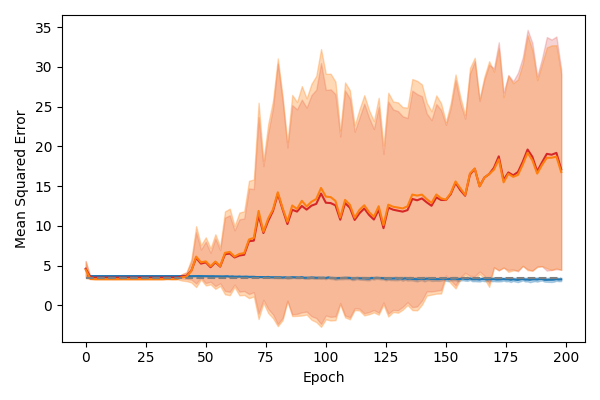}  
  \caption{$ A^{\pi^b}(s,a)$}
  \label{fig:rlstats6}
\end{subfigure}
\begin{subfigure}{.48\textwidth}
  \centering
  \includegraphics[width=\linewidth]{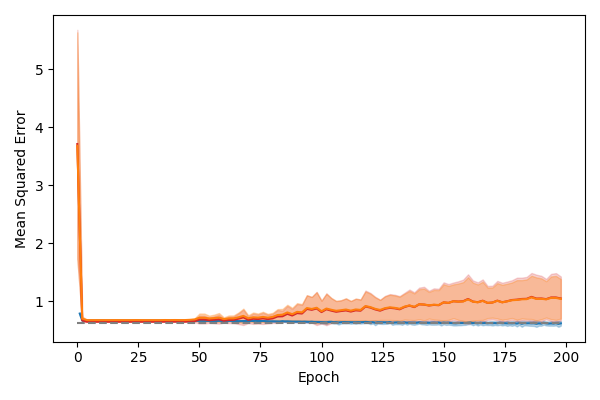}  
  \caption{$ S(s,a)$}
  \label{fig:rlstats7}
\end{subfigure}
 \caption{Loss profiles for training with other task statistics}
\label{fig:other_rlstats}
\vspace{-0.5cm}
\end{figure}

For training on these task statistics, cross entropy loss is used for optimality classification, mean square error is used for all other task statistics, and loss of the auxiliary task is additionally used for all task statistics for consistency. As shown in Fig.~\ref{fig:other_rlstats}, the models trained on these task statistics all tend to overfit: as soon as the training loss starts to decrease, the test and validation loss start to increase, both never decreasing below the baseline performance of predicting the label's mean. 

We speculate that modeling these task statistics is difficult due to \textit{time-aliasing} in the training data, in which two adjacent training inputs in two consecutive timesteps are very similar but have different labels determined by the timestep's task statistics. Such time-aliasing is less of a problem when modeling only non-zero reward categories since non-zero reward events are often separated by zero-reward steps. An important direction for future research is to find a mechanism to directly address the time-aliasing in data labels. 

We've also used a discount factor in computing the task statistics, treating \textit{Robotaxi} as an infinite-horizon MDP while the actual episodes have a finite trajectory length of 200 time steps. This could be another factor that influences our modeling of these task statistics. 

\section{Evaluating Robotic Sorting Task} \label{RoboticEval}

\begin{figure}[h]
    \centering
    \includegraphics[width=0.6\textwidth]{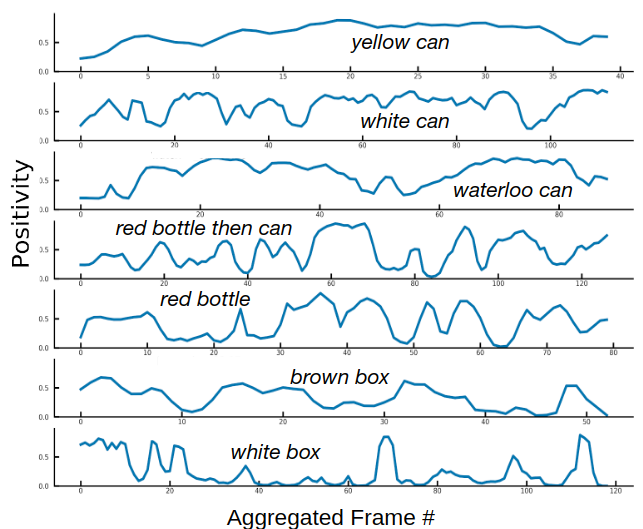}
    \caption{Sample plot of trajectory positivity score over aggregated frames } 
    \label{fig:sample_traj_score}
\end{figure}

\begin{figure}[h]
    \vspace{-0.1cm}
   \centering
    \includegraphics[width=0.4\textwidth]{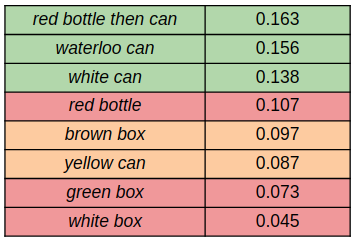}
    \vspace{-0.1cm}
    \captionof{figure}{Overall trajectory ranking by mean positivity score across subjects (each entry is colored by the trajectory's final return: green for $+2$, yellow for $0$, and red for $-1$) }
    \label{fig:overall_rank}
\end{figure}

To evaluate the robotic sorting trajectories. We consider each trajectory as an extended action and assume facial reactions along the trajectory are generated by a single latent state that represents the human's internal model of the robot's action. Therefore, the mean of the positivity score along each trajectory is used as the overall scoring of the trajectory. Fig.~\ref{fig:sample_traj_score} shows the positivity score along all 7 trajectories (episodes) of the robotics task from a participant. For each trajectory, the mean positivity score across all participants is then computed. Fig.~\ref{fig:overall_rank} shows the trajectory ranking related to all sorted items with their corresponding mean positivity score.

\clearpage